\documentclass[english,twocolumn,superscriptaddress]{revtex4-1}
\usepackage[T1]{fontenc}
\usepackage[latin9]{inputenc}
\usepackage{color}
\usepackage{array}
\usepackage{multirow}
\usepackage{amsmath}
\usepackage{amssymb}
\usepackage{graphicx}

\makeatletter

\providecommand{\tabularnewline}{\\}

\@ifundefined{textcolor}{}
{%
 \definecolor{BLACK}{gray}{0}
 \definecolor{WHITE}{gray}{1}
 \definecolor{RED}{rgb}{1,0,0}
 \definecolor{GREEN}{rgb}{0,1,0}
 \definecolor{BLUE}{rgb}{0,0,1}
 \definecolor{CYAN}{cmyk}{1,0,0,0}
 \definecolor{MAGENTA}{cmyk}{0,1,0,0}
 \definecolor{YELLOW}{cmyk}{0,0,1,0}
}

\makeatother

\usepackage{babel}
\begin{document}

\title{Criteria for genuine $N$-partite continuous-variable entanglement
and Einstein-Podolsky-Rosen steering}

\author{R.Y. Teh and M.D. Reid}

\email{rteh@swin.edu.au}

\email{mdreid@swin.edu.au}

\affiliation{Centre for Quantum Atom Optics, Swinburne University of Technology,
Melbourne, 3122 Australia }
\begin{abstract}
Following previous work, we distinguish between genuine $N$-partite
entanglement and full $N$-partite inseparability. Accordingly, we
derive criteria to detect genuine multipartite entanglement using
continuous variable (position and momentum) measurements. Our criteria
are similar but different to those based on the van Loock-Furusawa
inequalities, which detect full $N$-partite inseparability. We explain
how the criteria can be used to detect the genuine $N$-partite entanglement
of continuous variable states generated from squeezed and vacuum state
inputs, including the continuous variable Greenberger-Horne-Zeilinger
state, with explicit predictions for up to $N=9$$ $. This makes
our work accessible to experiment. For $N=3$, we also present criteria
for tripartite Einstein-Podolsky-Rosen (EPR) steering. These criteria
provide a means to demonstrate a genuine three-party EPR paradox,
in which any single party is steerable by the remaining two parties.

\end{abstract}
\maketitle

\section{Introduction}

There has been strong motivation to create and detect quantum states
that have many atoms \cite{wine}, photons, \cite{6dicke,svetexp,shalm-1}
or modes \cite{aokicv,seiji,threecolourcv} entangled. Beyond the
importance to the field of quantum information, such states provide
evidence for mesoscopic quantum mechanics \cite{ghz,seevinck,svetlichny,collins localnonlocal-2}.
In any such experiment, it is essential that one can clearly distinguish
the genuine $N$-partite entanglement of $N$ systems from the entanglement
produced by mixing quantum states with fewer than $N$ systems entangled.

Three systems labeled $1$, $2$, and $3$ are said to be \emph{genuinely
tripartite entangled} iff the density operator for the tripartite
system cannot be represented in the biseparable form \cite{bancalgendient,guh4}
\begin{eqnarray}
\rho_{BS} & = & P_{1}\sum_{R}\eta_{R}^{(1)}\rho_{23}^{R}\rho_{1}^{R}+P_{2}\sum_{R'}\eta_{R'}^{(2)}\rho_{13}^{R'}\rho_{2}^{R'}\nonumber \\
 &  & +P_{3}\sum_{R''}\eta_{R''}^{(3)}\rho_{12}^{R''}\rho_{3}^{R''},\label{eq:mixgen}
\end{eqnarray}
where $\sum_{k=1}^{3}P_{k}=1$ and $\sum_{R}\eta_{R}^{(k)}=1$. Here,
$\rho_{k}^{R}$ is an arbitrary quantum density operator for the system
$k$, while $\rho_{mn}^{R}$ is an arbitrary quantum density operator
for the two systems $m$ and $n$ ($k,m,n$ $\in\{1,2,3\}$). Thus,
for a system described by the biseparable state $\rho_{mn}^{R}\rho_{k}^{R}$,
the systems $m$ and $n$ can be bipartite entangled, but there is
no entanglement between $m$ and $k$, or $n$ and $k$. Similarly,
$N$ parties are ``genuinely $N$-partite entangled'' if all the
possible biseparable mixtures describing the $N$ parties are negated.

In this paper, we use the above definition to derive criteria sufficient
to confirm the genuine $N$-partite entanglement of $N$ systems,
as detected by continuous variable (CV) measurements, i.e., measurements
of position and momentum, or quadrature phase amplitudes. An application
of the criteria would be to witness the genuine entanglement of $N$
spatially separated optical field modes \cite{aokicv,seiji,threecolourcv}. 

The continuous variable (CV) case is an important one \cite{cv gaussrmp,rmp-1,slbvl rmp,eisertplenio}.
CV entanglement has significant applications to quantum information
technology, providing efficient deterministic teleportation \cite{tele}
and secure communication \cite{cry}. Moreover, CV entanglement can
give efficiently detected Einstein-Podolsky-Rosen correlations \cite{rmp-1,ou epr-1}
and evidence of the entanglement of multiple macroscopic systems,
consisting of many photons \cite{cv large}. The CV criteria can also
be applied to optomechanics, as a means to demonstrate the entanglement
of three or more mechanical harmonic oscillators \cite{optmehc}.

In order to claim genuine multipartite entanglement, it is necessary
to falsify all mixtures of the bipartitions as in Eq. (\ref{eq:mixgen}),
as opposed to negating that the system can be in any single one of
them. As pointed out by Shalm et al. \cite{shalm-1}, this leads to
two definitions $ $$-$ \emph{genuine $N$-paritite entanglement}
and \emph{full $N$-partite inseparability}$ $$-$ that have often
been used interchangeably in the literature but in fact mean different
things. This distinction for Gaussian states was also made by Hyllus
and Eisert \cite{hyllus and eisert}. In realistic experimental scenarios
where one cannot assume pure states, the task of detecting genuine
continuous variable (CV) multipartite entanglement poses a greater
challenge than detecting full multipartite inseparability. This means
that detecting genuine tripartite entanglement in the CV regime is
more difficult than has often been supposed. Most CV criteria that
have been applied to experiments assume Gaussian states \cite{Gaussian gencv,gausadesso},
or else do not in fact negate all \emph{mixtures} of bipartitions
(\ref{eq:mixgen}), and thus detect full multipartite inseparability,
rather than genuine multipartite entanglement \cite{cvsig,aokicv,seiji,threecolourcv}. 

One exception is the work of Shalm et al. \cite{shalm-1}. These authors
derive new CV criteria involving position and momentum observables.
Shalm et al. then adapt the criteria, to demonstrate the genuine tripartite
entanglement of three spatially separated photons using energy-time
measurements. A second exception is Armstrong et al. \cite{seiji2},
who derive a different criterion that is used to confirm the genuine
CV tripartite entanglement of three optical modes. Also, the recent
work of He and Reid \cite{genepr} gives criteria for genuine tripartite
EPR steering, which is a special type of tripartite entanglement.

Here, we present criteria for the detection of CV multipartite entanglement.
The criteria can be applied to the CV Greenberger-Horne-Zeilinger
(GHZ) states \cite{braunghz} that have been generated in the experiments
of Aoki et al. \cite{aokicv}, or the similar multipartite Einstein-Podolsky-Rosen
(EPR) entangled states generated in the experiments of Armstrong et
al. \cite{seiji}. In Secs. II and III, we present the necessary background,
and in Secs. IV and V derive criteria for the tripartite $N=3$ case.
In Sec. VIII, we provide algorithms for arbitrary $N$, and give explicit
predictions for up to $N=9$ modes, for the multi-mode CV GHZ- and
EPR-type entangled states. The effect of transmission losses is also
analyzed, in Sec. VII. Our criteria are based on the assumption that
the quantum uncertainty relations for position and momentum apply
to the measurements made on each system, and are not restricted to
pure or mixed Gaussian states. 

In Sec. VI, we analyze and derive criteria for ``genuine tripartite
EPR steering'' \cite{genepr}. ``EPR steering'' is the form of
quantum nonlocality introduced by EPR in their paradox of 1935 \cite{epr,hw-1}.
The term ``steering'' was introduced by Schrodinger to describe
the nonlocality highlighted by the paradox. EPR steering and the EPR
paradox were realized for CV measurements in the experiment of Ou
et al. \cite{ou epr-1}, based on the predictions explained in Ref.
\cite{rmp-1}. In short, verification of steering amounts to a verification
of entanglement, in a scenario where not all of the experimentalists
can be trusted to carry out the measurements properly \cite{hw-1,cv trust}.
This is an important consideration in device-independent quantum cryptography
\cite{one-sidedcrytpt}. The criteria developed in this paper are
likely to be useful to multiparty quantum cryptography protocols,
such as quantum secret sharing \cite{secretshare}.

The inequalities that we use to detect genuine $N$-partite entanglement
are similar to the van Loock-Furusawa inequalities \cite{cvsig}.
The van Loock-Furusawa inequalities are widely used, but are designed
to test for full multipartite inseparability, rather then genuine
multipartite entanglement. However, we show that one of the van Loock-Furusawa
inequalities will suffice to detect genuine tripartite entanglement,
and that tripartite entanglement and steering can be detected for
sufficient violation of other van Loock-Furusawa inequalities that
are used together as a set. Our work extends beyond the $N=3$ case.
We prove in Section VIII a general approach for deriving entanglement
criteria based on summation of inequalities that can negate each pure
biseparable state. Further, we establish that the genuine $N$-partite
entanglement of CV GHZ and certain multipartite EPR states can be
detected using a single suitably-optimized inequality.

\section{Distinguishing between genuine $N$-partite entanglement and full
$N$-partite inseparability}

The aim of this paper is to derive inequalities based on the assumption
(\ref{eq:mixgen}) of the biseparable mixture, and the $N$-party
extensions. The violation of these inequalities will then demonstrate
genuine tripartite entanglement, and, in the $N$-party case, genuine
$N$-partite entanglement. First, we explain the difference between
\emph{genuine $N$-paritite entanglement} and \emph{full $N$-partite
inseparability. }

\begin{figure}
\includegraphics[width=0.8\columnwidth]{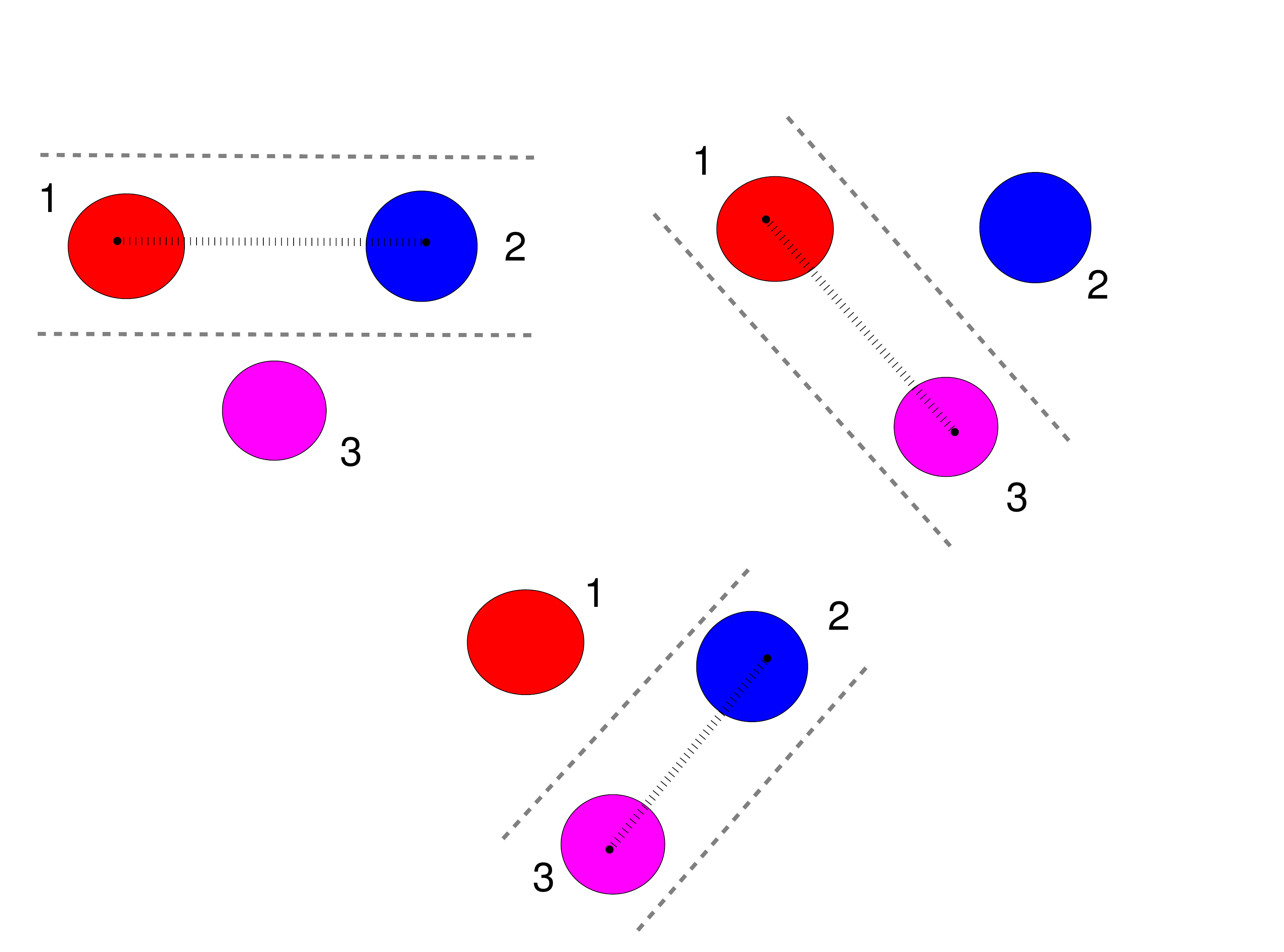}\protect\caption{(Color online) Schematic of the possible biseparable quantum states
for three systems, labeled $1,2,3$. Going clockwise from top left,
the biseparable states are $\rho_{12,3}$, $\rho_{13,2}$, and $\rho_{23,1}$
(associated with bipartitions $12-3$, $13-2$, and $23-1$, respectively).
\label{fig:Schematic-3}}
\end{figure}

We consider the three-party system described by
\begin{equation}
\rho_{km,n}=\sum_{R}\eta_{R}^{(n)}\rho_{km}^{R}\rho_{n}^{R}\label{eq:sepklm}
\end{equation}
where two but not three of the systems can be entangled. In this notation,
the $k,m,n$ denote three distinct systems, which in this paper will
be modes representing an optical field or a quantized harmonic oscillator.
The density operator $\rho_{km}^{R}$ can represent any quantum state
for the two modes $k$ and $m$, and can account for entanglement
between them. We denote the bipartition associated with the biseparable
density operator $\rho_{km}^{R}\rho_{n}^{R}$ (and $\rho_{km,n}$)
by $km-n$. The bipartitions for $N=3$ parties are depicted graphically
in Fig. \ref{fig:Schematic-3}. 

We suppose that each system is a single mode with boson operator $a_{j}$
($j=1,2,3$) and define the quadrature amplitudes as $x_{j}=(a_{j}+a_{j}^{\dagger})$
and $p_{j}=(a_{j}-a_{j}^{\dagger})/i$. Assuming the Heisenberg uncertainty
relation $\Delta x_{j}\Delta p_{j}\geq1$, the separability assumption
of (\ref{eq:sepklm}) implies the following sum and product inequalities:
\begin{equation}
(\Delta u)^{2}+(\Delta v)^{2}\geq2(|h_{n}g_{n}|+|h_{k}g_{k}+h_{m}g_{m}|)\label{eq:uvineq}
\end{equation}
and 
\begin{equation}
\Delta u\Delta v\geq|h_{n}g_{n}|+|h_{k}g_{k}+h_{m}g_{m}|,\label{eq:uvineq-1}
\end{equation}
where $u=h_{n}x_{n}+h_{k}x_{k}+h_{m}x_{m}$ and $v=g_{n}p_{n}+g_{k}p_{k}+g_{m}p_{m}$.
Here, $(\Delta x)^{2}$ denotes the variance of the quantum observable
$x$. The sum inequality was derived by van Loock and Furusawa \cite{cvsig}.
The product inequality is proved in the Appendix, and is stronger,
in that it will always imply the sum inequality (note the simple identity
$x^{2}+y^{2}\geq2xy$, that holds for any real numbers $x$ and $y$).

In their paper, van Loock and Furusawa consider the three inequalities
\begin{eqnarray}
B_{I}\equiv[\Delta(x_{1}-x_{2})]^{2}+[\Delta(p_{1}+p_{2}+g_{3}p_{3})]^{2} & \geq & 4,\nonumber \\
B_{II}\equiv[\Delta(x_{2}-x_{3})]^{2}+[\Delta(g_{1}p_{1}+p_{2}+p_{3})]^{2} & \geq & 4,\nonumber \\
B_{III}\equiv[\Delta(x_{1}-x_{3})]^{2}+[\Delta(p_{1}+g_{2}p_{2}+p_{3})]^{2} & \geq & 4,\nonumber \\
\label{eq:threeineq}
\end{eqnarray}
which are defined for arbitrary real parameters $g_{1}$, $g_{2}$,
and $g_{3}$. They point out, using Eq. (\ref{eq:uvineq}), that inequality
$B_{I}\geq4$ is implied by both the biseparable states $ $$\rho_{13,2}$
and $\rho_{23,1}$, which give separability between systems $1$ and
$2$. Similarly, the second inequality $B_{II}\geq4$ is implied by
the biseparable states $ $$\rho_{13,2}$ and $\rho_{12,3}$, while
the third inequality $B_{III}\geq4$ follows from biseparable states
$ $$\rho_{12,3}$ and $\rho_{23,1}$. 

In this way, van Loock and Furusawa show that the violation of any
two of the inequalities of Eq. (\ref{eq:threeineq}) is sufficient
to rule out that the system is described by any of the biseparable
states $\rho_{12,3}$, $\rho_{13,2}$, or $\rho_{23,1}$. This result
has been used in experimental scenarios \cite{aokicv,seiji} to give
evidence of a ``\emph{fully inseparable tripartite entangled state}''.
However, violating any two of the van Loock-Furusawa inequalities
is not\emph{ }in itself sufficient to confirm genuine tripartite entanglement,
as can be verified by the \emph{mixed} state example given in the
Appendix 4. The reason is that inequalities ruling out any of the
simpler cases of Eq. (\ref{eq:sepklm}) do not rule out the general
biseparable case of Eq. (\ref{eq:mixgen}) which considers mixtures
of the different bipartitions, $\rho_{12,3}$, $\rho_{13,2}$, or
$\rho_{23,1}$.

This point has been noted by Hyllus and Eisert \cite{hyllus and eisert}
and Shalm et al. \cite{shalm-1} and leads to two definitions in connection
with multipartite entanglement. For pure states, the two definitions
coincide, since a pure system cannot be in a mixture of states. For
experimental verification however, an unambiguous signature of genuine
tripartite entanglement becomes necessary, since one cannot assume
pure states.

Before continuing, it is useful to derive the \emph{product} version
of the van Loock-Furusawa inequalities, that are based on the product
uncertainty relation given by Eq. (\ref{eq:uvineq-1})\textcolor{black}{.
We define:}\textcolor{red}{{} }
\begin{eqnarray}
S_{I}\equiv\Delta(x_{1}-x_{2})\Delta(p_{1}+p_{2}+g_{3}p_{3}) & \geq & 2,\nonumber \\
S_{II}\equiv\Delta(x_{2}-x_{3})\Delta(g_{1}p_{1}+p_{2}+p_{3}) & \geq & 2,\nonumber \\
S_{III}\equiv\Delta(x_{1}-x_{3})\Delta(p_{1}+g_{2}p_{2}+p_{3}) & \geq & 2.\nonumber \\
\label{eq:threeineq-1}
\end{eqnarray}
In the Appendix, we show that the inequality $S_{I}\geq2$ is implied
by the biseparable states $ $$\rho_{13,2}$ and $\rho_{23,1}$. Similarly,
the second inequality $S_{II}\geq2$ is implied by the biseparable
states $ $$\rho_{13,2}$ and $\rho_{12,3}$, and the third inequality
$S_{III}\geq2$ by $ $$\rho_{12,3}$ and $\rho_{23,1}$. The product
versions are worth considering, given that the product uncertainty
relation, Eq. (\ref{eq:uvineq-1}), is stronger than the sum form,
Eq. (\ref{eq:uvineq}).

The van Loock-Furusawa approach is readily extended to tests of $N$-partite
full inseparability \cite{cvsig}. In that case, the possibility that
the system can be separable with respect to any of the possible bipartitions
is negated, by way of testing for violation of a set of inequalities.
However, generally, this does not eliminate the possibility that the
system could be in a mixture of biseparable states, that have only
($N-1$) or fewer modes entangled. Thus, stricter criteria are necessary
to confirm genuine $N$-partite entanglement.

\section{Genuine tripartite entangled states}

We are now motivated to derive criteria sufficient to prove genuine
tripartite entanglement, according to the definition of Eq. (\ref{eq:mixgen}).
Our criteria will be applied to two types of states known to be tripartite
entangled: the CV GHZ states and similar states, that we refer to
generally as CV EPR-type states.

\begin{figure}
\includegraphics[width=0.8\columnwidth]{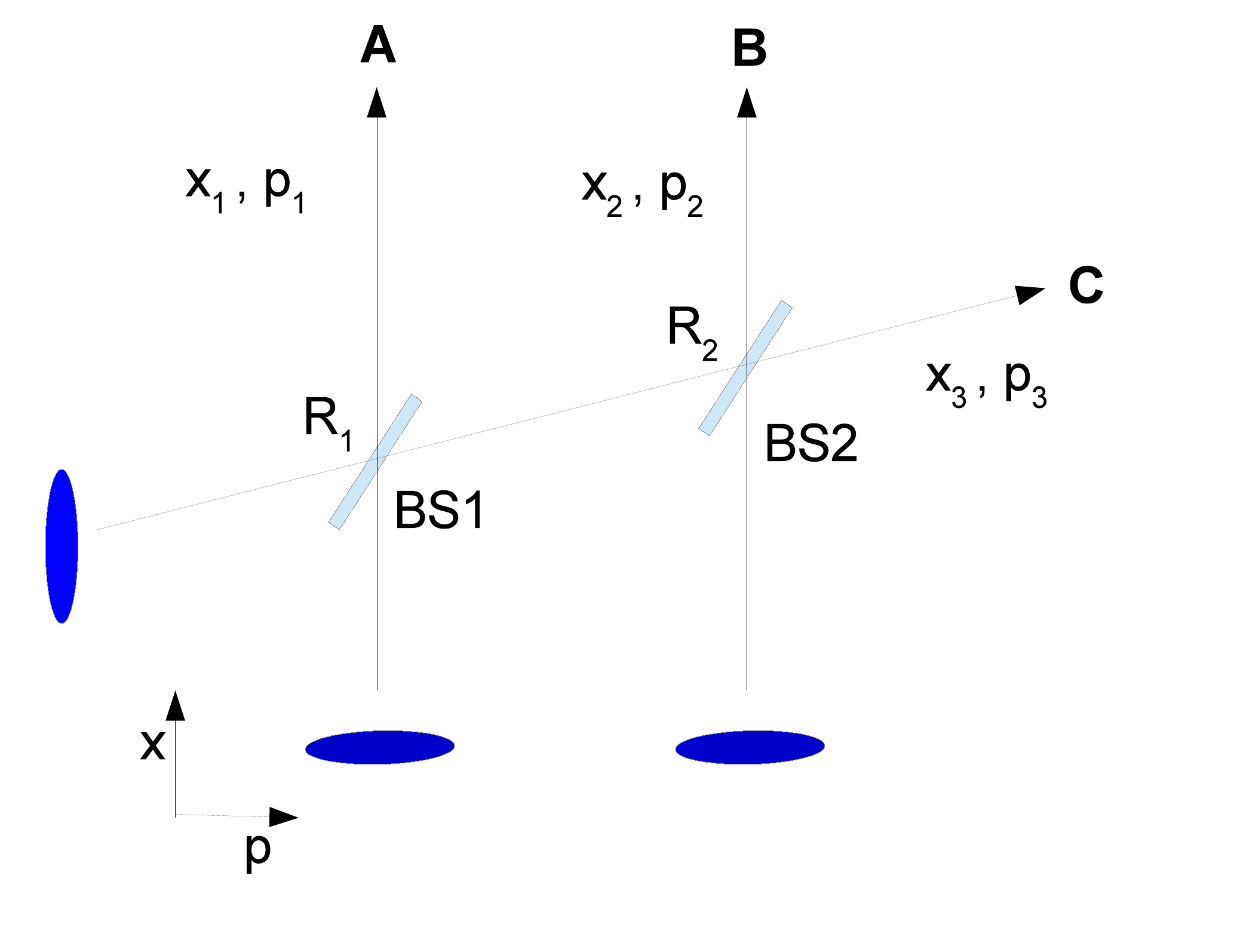}\protect\caption{(Color online) Schematic of the generation of a genuinely tripartite
entangled state, the CV GHZ state, using three squeezed input states.
Here, BS1 and BS2 symbolize beam splitters, with reflectivities given
by $R_{1}$ and $R_{2}$. For the GHZ state,\textcolor{red}{{} }\textcolor{black}{$R_{1}=\frac{1}{3}$
and $R_{2}=\frac{1}{2}$}. The $x_{j}$, $p_{j}$ refer to the two
orthogonal quadrature phase amplitudes of the optical mode $j$ ($j=1,2,3)$,
with boson operators $a_{j}$, $a_{j}^{\dagger}$. Three distinct
modes are formed at the three outputs, $A$, $B$, and \textcolor{black}{$C$.
The $x-p$ axis and ellipses depict the orientation of the squeezing
required.}\label{fig:Schematic}}
\end{figure}

The CV GHZ state \cite{braunghz} is generated using the configuration
shown in Fig. \ref{fig:Schematic} \cite{cvsig}. Two orthogonally
squeezed vacuum modes are the inputs of a beam splitter (BS1). This
creates a pair of entangled modes at the outputs of the first beam
splitter BS1. The entanglement is like that first discussed by EPR
in their argument for the completion of quantum mechanics, where the
positions and momenta (quadrature phase amplitudes) are both perfectly
correlated \cite{mdrepr,epr}. One of the entangled outputs is then
combined across a second beam splitter (BS2) using a third squeezed
state input. The squeeze parameters of the input states are assumed
equal, and of magnitude given by $r$. This means that in the idealised
experiment, each squeezed vacuum input has a quadrature variance given
by $\Delta x=e^{\mp r}$ and $\Delta p=e^{\pm r}$ (the sign depending
on the orientation of the squeezing and here we denote the ideal case
of pure squeezed inputs). More generally, the two entangled modes
could be created from parametric interactions \cite{mdrepr,heid two mode above threshold,twomodesq}
or similar atomic processes \cite{atomic tmss with dan}. Tripartite
entanglement can also be generated via three-photon parametric interactions
involving pump fields, as in the studies of Villar et al.. \cite{claude}.

A tripartite CV GHZ state is a simultaneous eigenstate of the position
difference $x_{i}-x_{j}$ ($i,j=1,2$, or $3$, $i\neq j$) and the
momentum sum $p_{1}+p_{2}+p_{3}$, and is formed in the limit of large
$r$. The experiment of Aoki et al. \cite{aokicv} used this generation
process to give an approximate realization of the CV GHZ state, to
the extent that they were able to demonstrate the full tripartite
inseparability of the three modes\textcolor{red}{{} }\textcolor{black}{(using
the van Loock-Furusawa inequalities of Eq. (\ref{eq:threeineq})).}
\begin{figure}
\includegraphics[width=0.8\columnwidth]{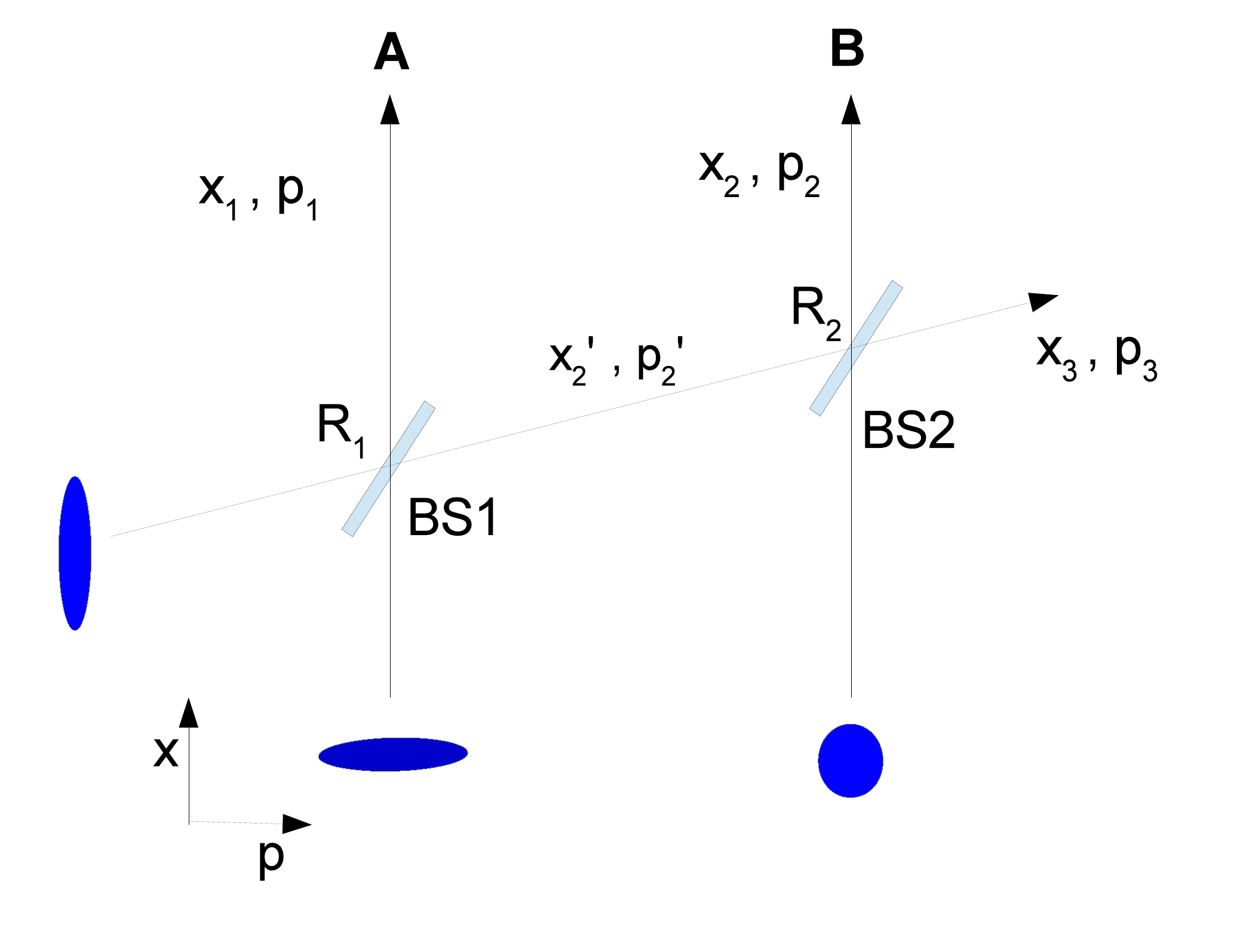}\protect\caption{(Color online) Schematic of the generation of a genuinely tripartite
entangled CV EPR-type state, using two squeezed input states and a
vacuum input at the second beam splitter. Here, BS1 and BS2 are beam
splitters, with reflectivities $R_{1}=R_{2}=0.5$. Labels as for Fig.
\ref{fig:Schematic}.\label{fig:Schematic-2} }
\end{figure}

In order to generate the second type of multipartite entangled state
(which we call the CV EPR-type state) the third squeezed input is
removed and replaced by a simple coherent vacuum state (Figure \ref{fig:Schematic-2}).
The multipartite entanglement of these sorts of states has been investigated
in the experiments of Armstrong et al. \cite{seiji,seiji2}. These
authors used the scheme of Figure \ref{fig:Schematic-2} and its $N$-party
extensions to generate states with a full $N$-partite inseparability,
up to $N=8$ modes. The van Loock-Furusawa inequality approach was
used to establish the inseparability.

The experimental confirmation of full $N$-partite inseparability
does not establish genuine $N$-partite entanglement, unless one can
justify pure states. In practice, this is not possible, because of
losses and the difficulty in achieving pure input squeezed states.
For this reason, we derive (in the following sections) criteria for
genuine $N$- partite entanglement, and then examine the effectiveness
of each criterion for the given CV states. We need to do this because
the criteria are \emph{sufficient}, but not \emph{necessary}, to detect
genuine multipartite entanglement. Calculations are therefore required
to determine which criterion should be used for a given CV state.
We will calculate the predictions for the criteria (which require
moments of the $x_{k}$ and $p_{k}$), using the simple unitary transformation
\begin{eqnarray}
a_{out,1} & = & \sqrt{R}a_{in,1}+\sqrt{1-R}a_{in,2}\nonumber \\
a_{out,2} & = & \sqrt{1-R}a_{in,1}-\sqrt{R}a_{in,2}\label{eq:bstrans}
\end{eqnarray}
that models the interaction of the modes at a beam splitter with reflectivity
$R$. Here, $a_{out,1}$, $a_{out,2}$ are the two output modes and
$a_{in,1}$, $a_{in,2}$ are the two input modes of the beam splitter.

\section{Criteria for Genuine $N$-partite entanglement: general approach }

We now explain a general method, that can be applied to detect the
$N$-partite entanglement. For a given $N$, the complete set of bipartitions
can be established. Let us suppose there are $X_{N}$ such bipartitions.
We index the bipartitions by $k=1,..,X_{N}$, and denote by $A_{k}$
and $B_{k}$ the two distinct sets of parties defined by the bipartition
$k$. For each bipartition $A_{k}-B_{k}$, we can establish an inequality
$I_{k}\geq4$ based on the assumption of separability of the system
density operator $\rho$ with respect to that bipartition, where the
$I_{k}$ is a sum $(\Delta u)^{2}+(\Delta v)^{2}$ of variances of
linear combinations $u$, $v$ of system observables $x_{j}$ and
$p_{j}$. This means that the observation of $I_{k}<4$ will imply
failure of separability (entanglement) between $A_{k}$ and $B_{k}$.
We can also establish similar inequalities $\mathcal{I}_{k}\geq2$
where $\mathcal{I}_{k}$ is a product $\Delta u\Delta v$.

We note that there will be many such inequalities for a given bipartition,
and that while $I_{k}<4$ suffices to imply inseparability between
$A_{k}$ and $B_{k}$, it is not necessary, so that the choice of
inequality is often intuitive, being dependent on the nature of the
quantum state. The van Loock-Furusawa inequalities are an example
of a set of inequalities $I_{k}$.

The violation of each of the inequalities $I_{k}\geq4$ ($k=1,...,X_{N})$
will not in itself imply genuine $N$-partite entanglement. However,
as might be expected, we can show that a large enough violation of
all the inequalities will in the end be sufficient. Thus, we establish
the following Result.

\textbf{Result (1):} Violation of the inequality 
\begin{equation}
\sum_{k=1}^{X_{N}}I_{k}\geq4\label{eq:result1}
\end{equation}
(or the inequality $\sum_{k=1}^{X_{N}}\mathcal{I}_{k}\geq2$ involving
the products) $ $is sufficient to imply $N$-partite genuine entanglement. 

\textbf{Proof: }We consider the $X_{N}$ bipartitions of the $N$-partite
system. We wish to negate the possibility that the system is described
by a mixture 
\begin{equation}
\rho_{BS}=\sum_{k=1}^{X_{N}}P_{k}\rho_{A_{k},B_{k}}\label{eq:n sep}
\end{equation}
where $P_{k}$ is a probability the system is separable across the
bipartition $ $$k$ (thus, $\sum_{k}P_{k}=1$). Separability across
the bipartition $k$ means that the density matrix is of the form
$\rho_{A_{k},B_{k}}=\sum_{R}\eta_{R}^{(k)}\rho_{A_{k}}\rho_{B_{k}}$,
where here $\rho_{A_{k}}$ and $\rho_{B_{k}}$ are density matrices
for subsystems $A_{k}$ and $B_{k}$ respectively. Consider a mixture
of states as given by a density operator $\rho=\sum_{R}P_{R}\rho_{R}$,
where $\sum_{R}P_{R}=1$ and $\rho_{R}$ is the density operator for
a component state. For any such mixture, the variance $(\Delta X)^{2}$
of an observable $X$ cannot be less than the weighted sum of the
variances of the component states: that is, 
\begin{equation}
(\Delta X)^{2}\geq\sum_{R}P_{R}(\Delta_{R}X)^{2}\label{eq:convesum}
\end{equation}
where $(\Delta_{R}X)^{2}$ denotes the variance of $X$ for the system
in the state $\rho_{R}$ \cite{hoftake}. Here, the observable $X$
is $u$ or $v$ as defined by Eqs. (\ref{eq:uvineq}) and (\ref{eq:uvineq-1}).
For two such observables, we have the result 
\begin{equation}
(\Delta X)^{2}+(\Delta Y)^{2}\geq\sum_{R}P_{R}\{(\Delta_{R}X)^{2}+(\Delta_{R}Y)^{2}\}\label{eq:sumconvex2}
\end{equation}
We can also prove a similar result for products of variances. In that
case, applying the Cauchy-Scwharz inequality, we can see that 
\begin{eqnarray}
(\Delta X)(\Delta Y) & \geq & \{\{\sum_{R}P_{R}(\Delta_{R}X)^{2}\}\{\sum_{R}P_{R}(\Delta_{R}Y)^{2}\}\}^{1/2}\nonumber \\
 & \geq & \sum_{R}P_{R}(\Delta_{R}X)(\Delta_{R}Y)\label{eq:convexprod}
\end{eqnarray}
Now, $I_{k}$ is the sum of variances. For example, $I_{k}=(\Delta u)^{2}+(\Delta v)^{2}$
can be the van Loock-Furusawa inequalities Eq. (\ref{eq:threeineq})
for certain values of $ $linear coefficients. Similarly, $\mathcal{I}_{k}=\Delta u\Delta v$
and can be the product inequalities Eq. (\ref{eq:threeineq-1}). If
the system is biseparable according to $\rho_{BS}$ of Eq. (\ref{eq:mixgen}),
then applying Eq. (\ref{eq:sumconvex2}) it follows that $ $
\[
I_{k}\geq\sum_{m=1}^{X_{N}}P_{m}I_{k,m}\geq4P_{k}
\]
where $I_{k,m}$ is the value of the sum of the variances that form
the expression $I_{k}$ evaluated over the biseparable state $\rho_{A_{m},B_{m}}$.
We have used that for the separable state $\rho_{A_{k},B_{k}}$, $ $
$I_{k}\geq4$. Summing over all $k$ and using that $\sum_{k=1}^{X_{N}}P_{k}=1$,
we obtain $\sum_{k}I_{k}\geq4$. Similarly, we can use Eq. (\ref{eq:convexprod})
to prove $\mathcal{I}_{k}\geq\sum_{m=1}^{X_{N}}P_{m}\mathcal{I}_{k,m}\geq2P_{k}$
and then that $\sum_{k}\mathcal{I}_{k}\geq2$. $\square$

Where there is a redundancy so that one of the inequalities $I_{k}\geq4$
is implied by more than one bipartition, we may be able to prove a
stronger criterion.  Certainly, if a \emph{single} inequality $I\geq4$
(or $\mathcal{I}\geq2$) can negate separability with respect to all
bipartitions $A_{k}-B_{k}$, then we can derive the following.

\textbf{Result (2):} Violation of the inequality 
\begin{equation}
I\geq4\label{eq:single}
\end{equation}
(or $\mathcal{I}\geq2$) which negates all of the biseparable states
$\rho_{A_{k},B_{k}}$ ($k=1,...,X_{N})$ is sufficient to imply $N$-partite
genuine entanglement. 

\textbf{Proof: }Consider a system described by the biseparable mixture
$\rho_{BS}$ of Eq. (\ref{eq:n sep}). Then using the results Eqs.
(\ref{eq:convesum}) and (\ref{eq:convexprod}) proved for mixtures,
it follows that for such a system 
\[
I\geq\sum_{m=1}^{X_{N}}P_{m}I_{k,m}\geq4
\]
where we have used the result that $I\geq4$ for every bipartition,
i.e. for every biseparable state $\rho_{A_{k},B_{k}}$ and hence that
each $I_{k.m}\geq4$. Similarly, $\mathcal{I}\geq2$.

The approach of using a single inequality is very valuable, once the
inequality can be identified. We will show how to use this method
for the CV GHZ and EPR-type states. Other criteria can be derived
where there are intermediate redundancies, as for the three van Loock-Furusawa
inequalities Eq. (\ref{eq:threeineq}). In that case, each inequality
will negate separability with respect to two bipartitions. We obtain
the following result.

\textbf{Criterion (1): }We confirm genuine tripartite entanglement,
if the following inequality is violated: 
\begin{equation}
B_{I}+B_{II}+B_{III}\geq8\label{eq:threesum-3}
\end{equation}
where $B_{I}\geq4$ , $B_{II}\geq4$ and $B_{III}\geq4$ are the van
Loock-Furusawa inequalities, Eq. (\ref{eq:threeineq}). We note that
$B_{I}$, $B_{II}$, $B_{III}$ is a function of the variable parameters
$g_{3}$, $g_{1}$, $g_{2}$ respectively.

\textbf{Proof: }For $N=3$ parties, there are three biseparable states
$\rho_{23,1}$, $\rho_{13,2}$, and $\rho_{12,3}$ that we index by
$k=1,2,3$ respectively. Consider any mixture of the form Eq. (\ref{eq:mixgen}),
which is Eq. (\ref{eq:n sep}) for $N=3$. Using the result Eq. (\ref{eq:convesum})
and the notation defined in the proof of Result (1), since $B_{I}$
is the sum of two variances, we can write 
\begin{eqnarray*}
B_{I} & \geq & P_{1}B_{I,1}+P_{2}B_{I,2}+P_{3}B_{I,3}\\
 & \geq & P_{1}B_{I,1}+P_{2}B_{I,2}\geq4(P_{1}+P_{2})
\end{eqnarray*}
This uses that we know the first two states of the mixture (for which
$k=1,2$) will satisfy the inequality, $B_{I}\geq4$. Hence, for any
mixture $B_{I}\geq4(P_{1}+P_{2})$. Similarly, $B_{II}\geq4(P_{2}+P_{3})$
and $B_{III}\geq4(P_{1}+P_{3})$. Then we see that since $\sum_{k=1}^{3}P_{k}=1$,
for any mixture it must be true that $B_{I}+B_{II}+B_{III}\geq8$.
$\square$

The product version of the criterion follows along similar lines.
The proof is similar to that of Criterion (1) and is given in the
Appendix.

\textbf{Criterion (2): }We confirm genuine tripartite entanglement
if the following inequality is violated: 
\begin{equation}
S_{I}+S_{II}+S_{III}\geq4\label{eq:threesum-1-2}
\end{equation}
where $S_{I}\geq2$ , $S_{II}\geq2$ and $S_{III}\geq2$ are the product
van Loock-Furusawa-type inequalities, Eq. (\ref{eq:threeineq-1}).

\section{Criteria for Genuine tripartite entanglement }

We now derive specific criteria to detect the genuine tripartite entanglement
of the tripartite entangled CV GHZ and EPR-type states.

\subsection{Criteria that use a single inequality}

First we examine the case where the criterion takes the form of a
single inequality involving just two variances, rather than the sum
of three inequalities, as in Eqs. (\ref{eq:threesum-3}) and (\ref{eq:threesum-1-2}).
Such criteria can be useful, but need to be tailored to the type of
tripartite entangled state. In this Section, we present several such
inequalities.

\textbf{\textcolor{black}{Criterion (3):}} The violation of the inequality
\begin{equation}
[\Delta(x_{1}-\frac{(x_{2}+x_{3})}{\sqrt{2}})]^{2}+[\Delta(p_{1}+\frac{(p_{2}+p_{3})}{\sqrt{2}})]^{2}\geq2\label{eq:crit1-5}
\end{equation}
is sufficient to confirm genuine tripartite entanglement.

\textbf{\textcolor{black}{Proof:}}\textcolor{red}{{} }Van Loock and
Furusawa showed that the inequality is satisfied by all three biseparable
states of types $\rho_{12,3}$, $\rho_{13,2}$, and $\rho_{23,1}$
\cite{cvsig}. Hence, the proof follows on using the Result (2), given
by Eq. (\ref{eq:single}). $\square$

Van Loock and Furusawa pointed out that this single inequality can
be used to negate all three separable bipartitions $12-3$, $13-2$,
and $23-1$, and hence to certify full tripartite inseparability.
However, the application of the Eq. (\ref{eq:convesum}) for mixtures
is needed to complete the proof that this single inequality is indeed
sufficient to certify genuine tripartite entanglement. Before continuing,
we write the product version of this criterion.

\textbf{\textcolor{black}{Criterion (4):}}\textcolor{black}{{} }The
violation of the inequality
\begin{equation}
\Delta(x_{1}-\frac{(x_{2}+x_{3})}{\sqrt{2}})\times\Delta(p_{1}+\frac{(p_{2}+p_{3})}{\sqrt{2}})\geq1\label{eq:crit1-1-1-1-1}
\end{equation}
is sufficient to confirm genuine tripartite entanglement. 

\textbf{\textcolor{black}{Proof}}\textcolor{black}{: }The uncertainty
relation $\Delta x_{j}\Delta p_{j}\geq1$ implies that the inequality
$\Delta u\Delta v\geq1$ holds for all three types of states $\rho_{12,3}$,
$\rho_{13,2}$, and $\rho_{23,1}$. This follows directly from the
result Eq. (\ref{eq:uvineq-1}). $\square$

We see immediately that violation of Eq. (\ref{eq:crit1-5}) will
always imply violation of Eq. (\ref{eq:crit1-1-1-1-1}). (Since $x^{2}+y^{2}\geq2xy$
for any two real numbers $x$, $y$). Thus, the product criterion
(4) is a stronger (better) criterion. However, where $\Delta(x_{1}-\frac{(x_{2}+x_{3})}{\sqrt{2}})=\Delta(p_{1}+\frac{(p_{2}+p_{3})}{\sqrt{2}})$,
the two criteria are mathematically equivalent. (We note $x^{2}+y^{2}=2xy$
iff $x=y$). This is the case for the states we consider in this paper,
but is not true in general. For some other states, entanglement criteria
based on products of variances have proved useful \cite{product,prod2}.

\begin{figure}
\includegraphics[width=1\columnwidth]{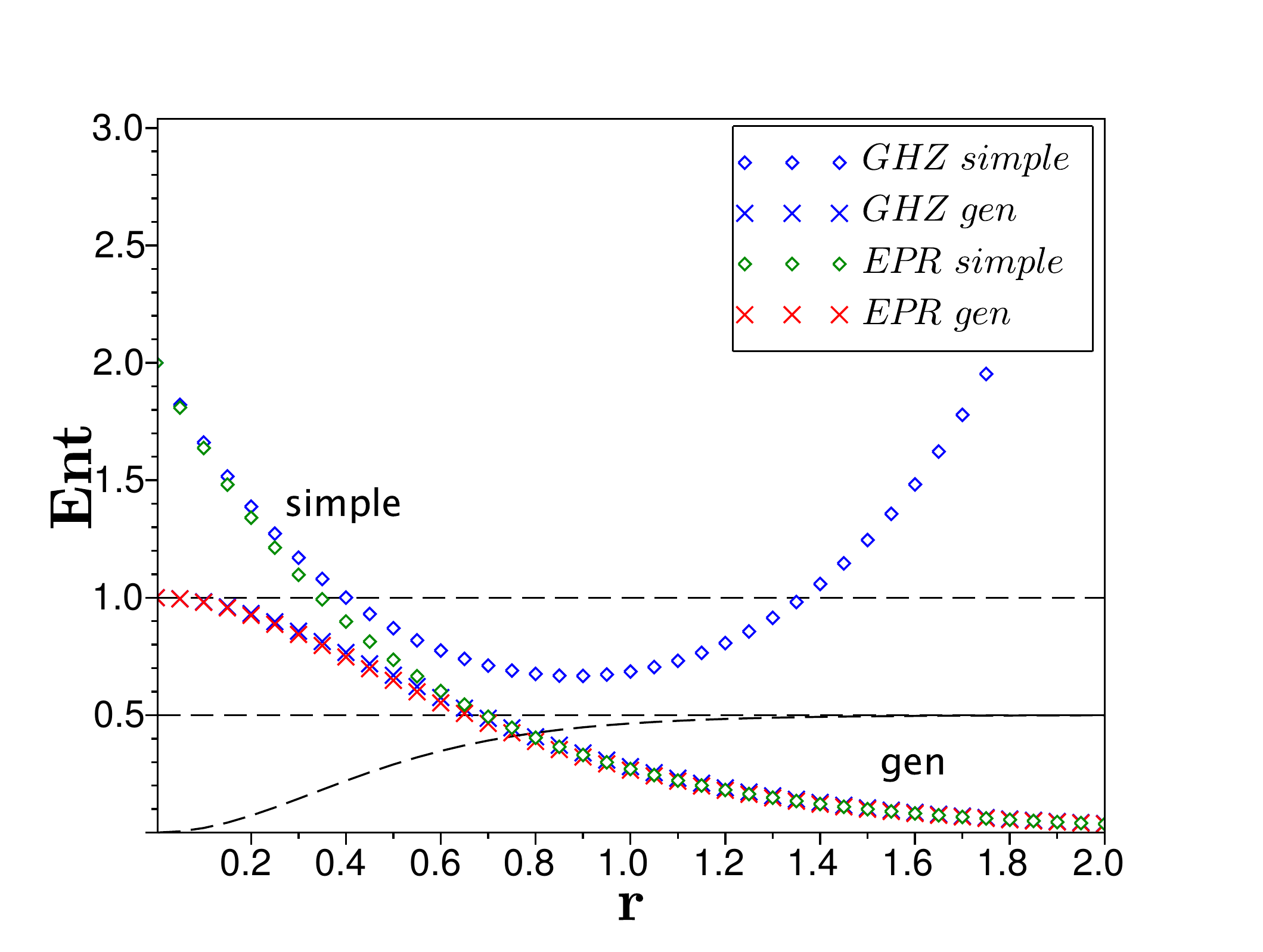}

\protect\caption{(Color online)\emph{ Detecting the genuine tripartite entanglement
of the CV GHZ and EPR-type states of Figures 2 and 3: }$Ent<1$ signifies
genuine tripartite entanglement. Here, $r$ is the squeezing parameter
of the input states shown in Figs. \ref{fig:Schematic}and\ref{fig:Schematic-2}.
The curves labeled ``simple'' are for the (top blue) GHZ and (second
green) EPR-type states, using the simple criteria (\ref{eq:crit1-5})
and (\ref{eq:crit1-1-1-1-1}) (which give indistinguishable results).
The two lower curves labeled ``gen'' are for the GHZ and EPR-type
states, using the generalized criteria (\ref{eq:big}) and (\ref{eq:prodgen-1})
(which give indistinguishable results). Here, $Ent=\frac{(\Delta u)^{2}+(\Delta v)^{2}}{2}$
and $Ent=\Delta u\Delta v$ for the criteria involving sums and products
respectively, where $u=x_{1}+h(x_{2}+x_{3})$, $v=p_{1}+g(p_{2}+p_{3})$.
The choices for $h$ and $g$ are given in Table \ref{tableBS-1}
for the generalized criteria, and are $g=-h=\frac{1}{\sqrt{2}}$ for
the simple criteria. Genuine tripartite steering is signified when
$Ent$ drops below the black dashed line\textcolor{red}{{} }for the
``gen'' case, and below $0.5$ for the ``simple'' curves. All
curves except the ``simple GHZ'' become indistinguishable at larger
$r$. \label{fig:gentri3-2-3}\textcolor{red}{}}
\end{figure}

The two simple criteria (\ref{eq:crit1-5}) and (\ref{eq:crit1-1-1-1-1})
are effective for demonstrating the genuine tripartite entanglement
of the EPR-type state, as shown by in the recent paper of Armstrong
et al. \cite{seiji2} where the product criterion (\ref{eq:crit1-5})
was derived. The predictions are plotted in Figure \ref{fig:gentri3-2-3}. 

For the CV GHZ state, it is better to consider a more generalised
criterion that allows arbitrary coefficients.

\textbf{\textcolor{black}{Criterion (5):}}\textcolor{black}{{} }Violation
of the inequality
\begin{eqnarray}
(\Delta u)^{2}+(\Delta v)^{2} & \geq & 2\min{\{|g_{3}h_{3}|+|h_{1}g_{1}+h_{2}g_{2}|,}\nonumber \\
 &  & |g_{2}h_{2}|+|h_{1}g_{1}+h_{3}g_{3}|,\nonumber \\
 &  & |g_{1}h_{1}|+|g_{2}h_{2}+h_{3}g_{3}|\}\label{eq:big}
\end{eqnarray}
where we define $u=h_{1}x_{1}+h_{2}x_{2}+h_{3}x_{3}$ and $v=g_{1}p_{1}+g_{2}p_{2}+g_{3}p_{3}$
is sufficient to confirm genuine triparite entanglement. Here, $g_{i}$,
$h_{i}$ are real constants ($i=1,2,3$).

\textbf{\textcolor{black}{Proof: }}Using Eq. (\ref{eq:uvineq}), we
see that the bipartition $\rho_{12}\rho_{3}$ implies 
\[
(\Delta u)^{2}+(\Delta v)^{2}\geq2\{|g_{3}h_{3}|+|h_{1}g_{1}+h_{2}g_{2}|\},
\]
 the bipartition $\rho_{13}\rho_{2}$ implies 
\[
(\Delta u)^{2}+(\Delta v)^{2}\geq2\{|g_{2}h_{2}|+|h_{1}g_{1}+h_{3}g_{3}|\},
\]
 and the bipartition $\rho_{23}\rho_{1}$ implies 
\[
(\Delta u)^{2}+(\Delta v)^{2}\geq2\{|g_{1}h_{1}|+|g_{2}h_{2}+h_{3}g_{3}|\}.
\]
 Thus, using the relation Eq. (\ref{eq:convesum}), we see that any
mixture Eq. (\ref{eq:mixgen}) will imply Eq. (\ref{eq:big}). $\square$

\begin{table}[h]
\protect\caption{Values of $g$ and $h$ used for the plots of Fig \ref{fig:gentri3-2-3}.\label{tableBS-1}}

\begin{tabular}{|c|>{\centering}p{0.2\columnwidth}|>{\centering}p{0.2\columnwidth}|>{\centering}p{0.2\columnwidth}|>{\centering}p{0.15\columnwidth}|}
\hline 
\multirow{2}{*}{\textbf{r}} & \multicolumn{2}{c|}{\textbf{CV GHZ}} & \multicolumn{2}{c|}{\textbf{EPR}}\tabularnewline
 & \multicolumn{1}{>{\centering}p{0.2\columnwidth}}{g} & h & \multicolumn{1}{>{\centering}p{0.2\columnwidth}}{g} & h\tabularnewline
\hline 
0\textbf{ } & 0 & 0 & 0 & 0\tabularnewline
\hline 
0.25\textbf{ } & 0.36 & -0.27 & 0.33 & -0.33\tabularnewline
\hline 
0.5 & 0.68 & -0.40 & 0.54 & -0.54\tabularnewline
\hline 
0.75 & 0.86 & -0.46 & 0.64 & -0.64\tabularnewline
\hline 
1 & 0.95 & -0.49 & 0.68 & -0.68\tabularnewline
\hline 
1.5 & 0.99 & -0.50 & 0.70 & -0.70\tabularnewline
\hline 
2\textbf{ } & 1.00 & -0.50 & 0.70 & -0.70\tabularnewline
\hline 
\end{tabular}
\end{table}

The Criterion (5) is valid for any choice of coefficients $g_{i}$
and $h_{i}$, which are real constants. If the inequality is violated,
then the experimentalist can conclude the three modes are genuine
tripartite entangled. However, as the criteria are sufficient but
not necessary for entanglement, it cannot be assumed that the inequality
will be violated, even where there is entanglement present. In a practical
situation for a given entangled state, it is best to analyze in advance
the optimal values for $g$, $h$. These optimal values are defined
as giving the smallest ratio of the left to right side of the inequality,
for a given quantum state.

An optimization was carried out numerically, for a simpler version
of the inequalities obtained as follows: A simpler version of the
inequality (\ref{eq:big}) is obtained, if we select the values $g_{1}=h_{1}=1$,
$h_{2}=h_{3}=h$, and $g_{2}=g_{3}=g$ so that $u=x_{1}+h(x_{2}+x_{3})$
and $v=p_{1}+g(p_{2}+p_{3})$, and then restrict to $gh<0$ and $|gh|<1$.
We note that the right side of Eq. (\ref{eq:big}) becomes $2\min\{|gh|+|1-|gh||,1+2|gh|\}$.
Eq. (\ref{eq:big}) then takes the form 
\begin{equation}
(\Delta u)^{2}+(\Delta v)^{2}\geq2.\label{eq:geneprcrti-1simple-1}
\end{equation}
Violation of this inequality will confirm genuine tripartite entanglement
(as a special case of Criterion (5)). The theoretical prediction for
the optimal value of gain constants $g$ and $h$ was found rigorously
by a numerical search over all values. The optimized values and associated
violation of the inequalities for the CV GHZ and EPR-type states are
given in Table I and Fig. \ref{fig:gentri3-2-3}.

An experimental set-up to detect the genuine tripartite entanglement
is like that described in Ref. \cite{cvsig} and implemented in the
experiment \cite{aokicv}, to detect full tripartite inseparability.
Ideally, in a tripartite version of an EPR experiment, the quadrature
amplitudes would be measured simultaneously in a spacelike separated
way at each of the three locations \cite{epr,rmp-1}. The inequalities
are tested by direct insertion of the results into the inequality,
with the $g$ and $h$ serving as numbers. In the experiments modeled
after squeezing measurements \cite{aokicv,mdrepr,ou epr-1}, the final
variances are measured directly as noise levels, and the $g$ and
$h$ factors are introduced by classical gains in currents.

We also derive the product form of the generalized criterion (\ref{eq:big}).
\textcolor{black}{The proof is similar to that for Criterion Eq. (\ref{eq:big})
and is given in the Appendix.}

\textbf{\textcolor{black}{Criterion (6):}}\textcolor{black}{{} Genuine
tripartite entanglement is observed if the inequality 
\begin{eqnarray}
\Delta u\Delta v & \geq & \min{\{|g_{3}h_{3}|+|h_{1}g_{1}+h_{2}g_{2}|,}\nonumber \\
 &  & \{|g_{2}h_{2}|+|h_{1}g_{1}+h_{3}g_{3}|,\nonumber \\
 &  & |g_{1}h_{1}|+|g_{2}h_{2}+h_{3}g_{3}|\}\label{eq:prodgen-1}
\end{eqnarray}
is violated.} With the choice of values for $g_{i}$ and $h_{i}$
explained for the inequality (\ref{eq:geneprcrti-1simple-1}) and
as given in Table \ref{tableBS-1}, the inequality (\ref{eq:prodgen-1})
takes the simpler form 
\begin{equation}
\Delta u\Delta v\geq1.\label{eq:geneprcrti-1simple-1-1-1}
\end{equation}

While the optimal values of the coefficients $g$ and $h$ were found
by numerical search, it is possible to deduce these values from the
physics associated with the different entangled states, at least in
the limit of large $r$. We see from the results of Table \ref{tableBS-1}
and Figure \ref{fig:gentri3-2-3} that for larger $r$, the genuine
tripartite entanglement of the CV GHZ state is detected by violation
of the inequality 
\begin{equation}
[\Delta(x_{1}-\frac{(x_{2}+x_{3})}{2})]^{2}+[\Delta(p_{1}+p_{2}+p_{3})]^{2}\geq2\label{eq:crit1-2}
\end{equation}
This is to be expected, since the CV GHZ state formed in the limit
of large $r$ is by definition the simultaneous eigenstate of position
difference $x_{i}-x_{j}$ ($i,j=1,2$ or $3$, $i\neq j$) and the
momentum sum $p_{1}+p_{2}+p_{3}$.

Similarly, for the EPR-type states of Fig. \ref{fig:Schematic-2},
the simple Criterion of Eq. (\ref{eq:crit1-5}) \textcolor{black}{(and
Eq. (\ref{eq:crit1-1-1-1-1}))} is in fact optimal at large $r$.
This can be understood as follows \cite{cvsig}: The two entangled
modes labeled $1$ and $2'$ in Fig. 3 possess an EPR correlation
as $r\rightarrow\infty$, so that simultaneously, both $[\Delta(x_{1}-x'_{2})]^{2}\rightarrow0$
and $[\Delta(p_{1}+p'_{2})]^{2}\rightarrow0$ where $x'_{2}$ and
$p'_{2}$ are the quadratures of the mode defined as $2'$. On examining
the model of Eq. (\ref{eq:bstrans}) for the beam splitter interaction
$BS2$, we put $a_{out,1/2}=a_{2/3}$, $a_{in,1}=a_{2'}$ and $a_{in,2}=a_{vac}$
where $a_{vac}$ is the boson operator for the vacuum mode input to
$BS2$. Then we see that for $R=0.5$, $a_{2'}=\frac{1}{\sqrt{2}}(a_{2}+a_{3})$,
which leads to the solution $x'_{2}=\frac{1}{\sqrt{2}}(x_{2}+x_{3})$
and $p'_{2}=\frac{1}{\sqrt{2}}(p_{2}+p_{3})$.\textcolor{red}{{} }Thus,
the EPR correlation of the original beams $1$ and $2'$ is transformed
into a tripartite EPR correlation that satisfies the Criterion (3)
of Eq. (\ref{eq:crit1-5}). This is the reason why we call these states
``EPR-type''. We note that as the EPR (or GHZ) correlation increases
(as it does with large $r$), the associated variances reduce, so
the amount of violation of the inequalities gives an indication of
the strength of that type of EPR (GHZ) entanglement.

We point out that the noise reduction required to demonstrate the
genuine tripartite entanglement is considerable, in the sense of being
beyond that necessary to demonstrate simple quantum squeezing, or
bipartite entanglement. Let us consider the group of modes $\{2,3\}$
created at the output of the second beam splitter $BS2$ as shown
in Figure \ref{fig:Schematic-2}. Bipartite entanglement between mode
$1$ and the combined group of modes $\{2,3\}$ can be certified when
$(\Delta u)^{2}+(\Delta v)^{2}<4$, which corresponds to a noise reduction
below the noise level of the quantum vacuum (measured by $4$ in this
case). The bipartite entanglement condition can be verified using
the techniques of Refs. \cite{duan-1,product}. Thus, the Criterion
(3) of Eq. (\ref{eq:crit1-5}) to confirm genuine tripartite entanglement
requires 50\% greater violation than to confirm ordinary bipartite
entanglement.\textbf{}

\subsection{Criteria using van Loock-Furusawa inequalities }

\begin{figure}
\includegraphics[width=1\columnwidth]{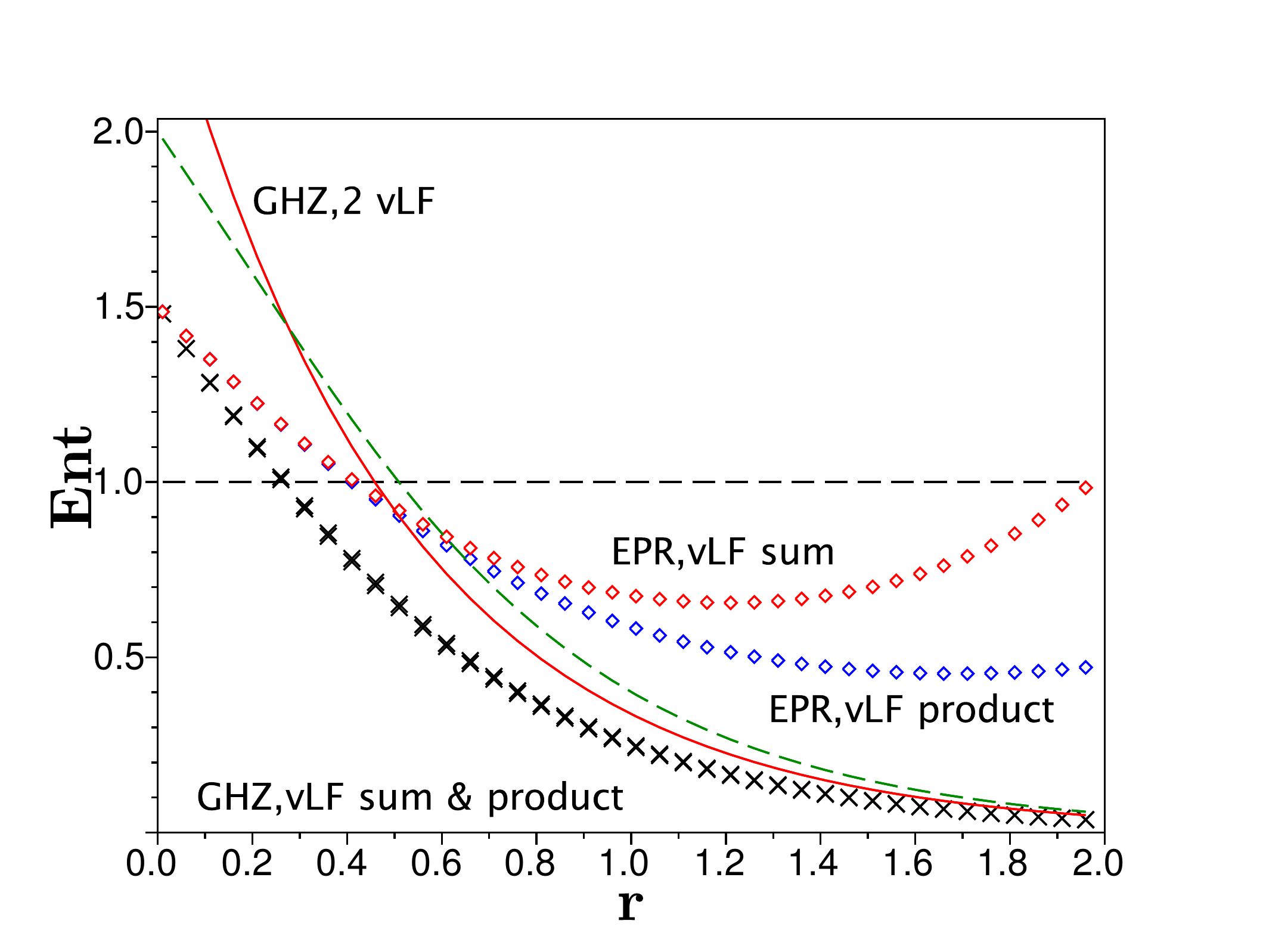}

\protect\caption{(Color online)\emph{ Detecting genuine $N$-partite entanglement by
summation of the violation of van Loock-Furusawa (vLF) inequalities,
or their product versions. }$Ent<1$ signifies genuine tripartite
entanglement; $Ent<0.5$ signifies genuine tripartite steering. The
description of states and the meaning of $r$ is as in Figure 4. For
the Criterion (4) given by Eq. (15), $Ent=(S_{I}+S_{II}+S_{III})/4$
(product version) and for the Criterion (3) given by Eq. (14), $Ent=(B_{I}+B_{II}+B_{III})/8$
(sum version). The choice of $g_{i}$'s is given in Table \ref{tableBS-1-2}.
The black (lower) crosses and blue (lower) diamonds give results for
the product criterion, for GHZ and EPR-type states, respectively,
with $N=3$. The Criteria (3) and (4) give indistinguishable results
for the GHZ state.\textcolor{red}{{} }The\textbf{ }upper red diamond
curve is the Criterion (3) for the $N=3$ EPR-type state. The red
line gives Criterion (7) (Eq. (\ref{eq:critasym})), involving just
two vLF inequalities, for the GHZ state, where $Ent=(B_{I}+B_{II})/4$.\textcolor{red}{{}
}The green dashed line is the Criterion (9) (Eq. (\ref{eq:four}))
for the GHZ state ($N=4$).\label{fig:gentri3-1}}
\end{figure}

Violation of the van Loock-Furusawa inequalities (Eq. (5)) have been
measured or calculated in numerous situations (including \cite{claude,aokicv,murraytri,seiji}).
In Figure \ref{fig:gentri3-1}, we use the Criteria (1) and (2), as
given by Eqs. (\ref{eq:threesum-3} and \ref{eq:threesum-1-2}), to
show that it is possible to verify the genuine tripartite entanglement
using the van Loock-Furusawa inequalities, provided there is enough
violation of the inequalities.

For symmetric systems such as the CV GHZ state, where $B_{I}=B_{II}=B_{III}$,
the condition (\ref{eq:threeineq-1}) of Criterion (1) requires $B_{I}<8/3$.
This level of noise reduction (which is $2/3$ the vacuum noise level)
would seem feasible in the set-up of experiment \cite{aokicv}. The
ideal CV GHZ state clearly violates the inequality, since in that
case $B_{I}=B_{II}=B_{III}\rightarrow0$ as $r\rightarrow\infty$.
The inequality for $g_{i}=1$ has been derived by Shalm et al.. \cite{shalm-1}.
We note from Table \ref{tableBS-1-2} that for the GHZ state the values
of $g_{1}=g_{2}=g_{3}=1$ are indeed optimal as $r\rightarrow\infty$.
The criterion derived here is valid for arbitrary $g$'s, which we
see from Table \ref{tableBS-1-2} is useful for the EPR-type states
of Fig. \ref{fig:Schematic-2}. These EPR-type states do not have
symmetry with respect to all three modes.

The effectiveness of the criteria is shown in the Figure \ref{fig:gentri3-1}
for the CV GHZ and EPR-type states. It is not surprising that the
criteria are more effective in the case of the GHZ states. This is
because the van Loock-Furusawa inequalities include terms involving
the variance of $x_{k}-x_{m}$ ($k\neq m$) which for the GHZ state
(but not the EPR-type state) will be small as $r\rightarrow\infty$. 

\begin{table}[h]
\protect\caption{Values of $g_{i}$ ($i=1,2,3$) for the plots of Figure \ref{fig:gentri3-1}.
The same values are used for the sum and product versions of the criteria.
\label{tableBS-1-2}}

\begin{tabular}{|>{\centering}p{0.1\columnwidth}|>{\centering}p{0.12\columnwidth}|>{\centering}p{0.13\columnwidth}|>{\centering}p{0.12\columnwidth}|>{\centering}p{0.12\columnwidth}|>{\centering}p{0.12\columnwidth}|>{\centering}p{0.12\columnwidth}|}
\hline 
r & \multicolumn{1}{>{\centering}p{0.12\columnwidth}}{~

\textbf{$g_{1}$}} & \multicolumn{1}{>{\centering}p{0.13\columnwidth}}{\textbf{GHZ}

\textbf{$g_{2}$}} & ~

\textbf{$g_{3}$} & \multicolumn{1}{>{\centering}p{0.12\columnwidth}}{~

\textbf{$g_{1}$}} & \multicolumn{1}{>{\centering}p{0.12\columnwidth}}{\textbf{EPR}

\textbf{$g_{2}$}} & ~

\textbf{$g_{3}$}\tabularnewline
\hline 
\ 0\textbf{ } & 0 & 0 & 0 & 0 & 0 & 0\tabularnewline
\hline 
0.25\textbf{ } & 0.53 & 0.53 & 0.53 & 0.63 & 0.29 & 0.29\tabularnewline
\hline 
0.5 & 0.81 & 0.81 & 0.81 & 1.08 & 0.44 & 0.44\tabularnewline
\hline 
0.75 & 0.93 & 0.93 & 0.93 & 1.28 & 0.50 & 0.50\tabularnewline
\hline 
1 & 0.97 & 0.97 & 0.97 & 1.36 & 0.50 & 0.50\tabularnewline
\hline 
1.5 & 1.00 & 1.00 & 1.00 & 1.41 & 0.46 & 0.46\tabularnewline
\hline 
2\textbf{ } & 1.00 & 1.00 & 1.00 & 1.41 & 0.43 & 0.43\tabularnewline
\hline 
\end{tabular}
\end{table}

\subsection{Criteria involving just two van Loock-Furusawa inequalities}

The following criterion involving just two inequalities but with $g_{i}=1$
has been derived by Shalm et al. \cite{shalm-1}.

\textbf{\textcolor{black}{Criterion (7):}}\textbf{ }We can confirm
genuine tripartite entanglement, if any two of the inequalities $B_{I}\geq4$,
$B_{II}\geq4$, $B_{III}\geq4$ given by Eq. (\ref{eq:threeineq})
with $g_{1}=g_{2}=g_{3}=1$ are violated by a sufficient margin, so
that 
\begin{equation}
B_{I}+B_{II}<4\label{eq:critasym}
\end{equation}
 (or $B_{I}+B_{III}<4$, or $B_{II}+B_{III}<4$).

The symmetry of the GHZ state means that the genuine tripartite entanglement
is detected using any one of these the inequalities. Where losses
are important, this can change. These criteria are not effective in
detecting the genuine tripartite entanglement of the EPR-type states,
for the reasons discussed above, that the variances of the van Loock-Furusawa
inequalities do not capture the correlated observables in this case.

\section{Criteria for genuine tripartite EPR steering}

We now consider criteria to detect the type of entanglement called
``genuine tripartite EPR steering''. EPR steering is a nonlocality
associated with the EPR paradox, that can be regarded in some sense
intermediate between entanglement and Bell's nonlocality \cite{hw-1,bell}.
We follow and expand on the methods of Ref. \cite{genepr}. The criteria
are the same inequalities as before, but with stricter bounds. The
physical significance of EPR steering is that it allows detection
of the entanglement even when some of the parties or measurement devices
associated with the systems $i=1,2,3$ cannot be trusted \cite{cv trust,one-sidedcrytpt}.
For example, we may not be able to assume that the results reported
by some parties are actually the result of quantum measurements $\hat{x}$
or $\hat{p}$. This can be important where the entanglement is used
for quantum key distribution \cite{one-sidedcrytpt}. 

Consider three measurements $X_{1}$, $X_{2}$ and $X_{3}$ made on
each of three distinct systems (also referred to as parties). Where
the composite system is given by the biseparable density matrix $\rho_{BS}$
of Eq. (\ref{eq:mixgen}), we note that any average $\langle X_{1}X_{2}X_{3}\rangle$
is expressible as

\begin{eqnarray}
\langle X_{1}X_{2}X_{3}\rangle & = & P_{1}\sum_{R}\eta_{R}^{(1)}\langle X_{2}X_{3}\rangle_{R}\langle X_{1}\rangle_{R,\rho}\nonumber \\
 &  & +P_{2}\sum_{R}\eta_{R}^{(2)}\langle X_{1}X_{3}\rangle_{R}\langle X_{2}\rangle_{R,\rho}\nonumber \\
 &  & +P_{3}\sum_{R}\eta_{R}^{(3)}\langle X_{1}X_{2}\rangle_{R}\langle X_{3}\rangle_{R,\rho}\label{eq:lhs-1}
\end{eqnarray}
Here all averages $\langle\rangle$ are those of a quantum density
matrix, and the $\rho$ subscript reminds us of that. To signify genuine
tripartite Bell nonlocality \cite{bell}, however, one needs to falsify
a stronger assumption. This can be done, if we falsify (\ref{eq:lhs-1}),
but without the assumption that the averages $\langle\rangle$ are
necessarily those of quantum states: They can be averages for hidden
variable states, as defined by Bell and Svetlichny \cite{svetlichny,gallego }. 

To signify \emph{genuine tripartite steering} \cite{genepr}, it is
sufficient to falsify a \emph{hybrid} local-nonlocal ``biseparable
Local Hidden State (LHS) model'', which is a multiparty extension
of the bipartite LHS models defined in Refs. \cite{hw-1,ericmultisteer}.
In that case, the averages $\langle X_{k}X_{m}\rangle_{R}$ (that
are without the subscript $\rho$) can be hidden variable averages,
whereas those for the single system $\langle X_{n}\rangle_{R,\rho}$
(written with the subscript $\rho$) are quantum averages. 

\begin{figure}
\includegraphics[width=0.8\columnwidth]{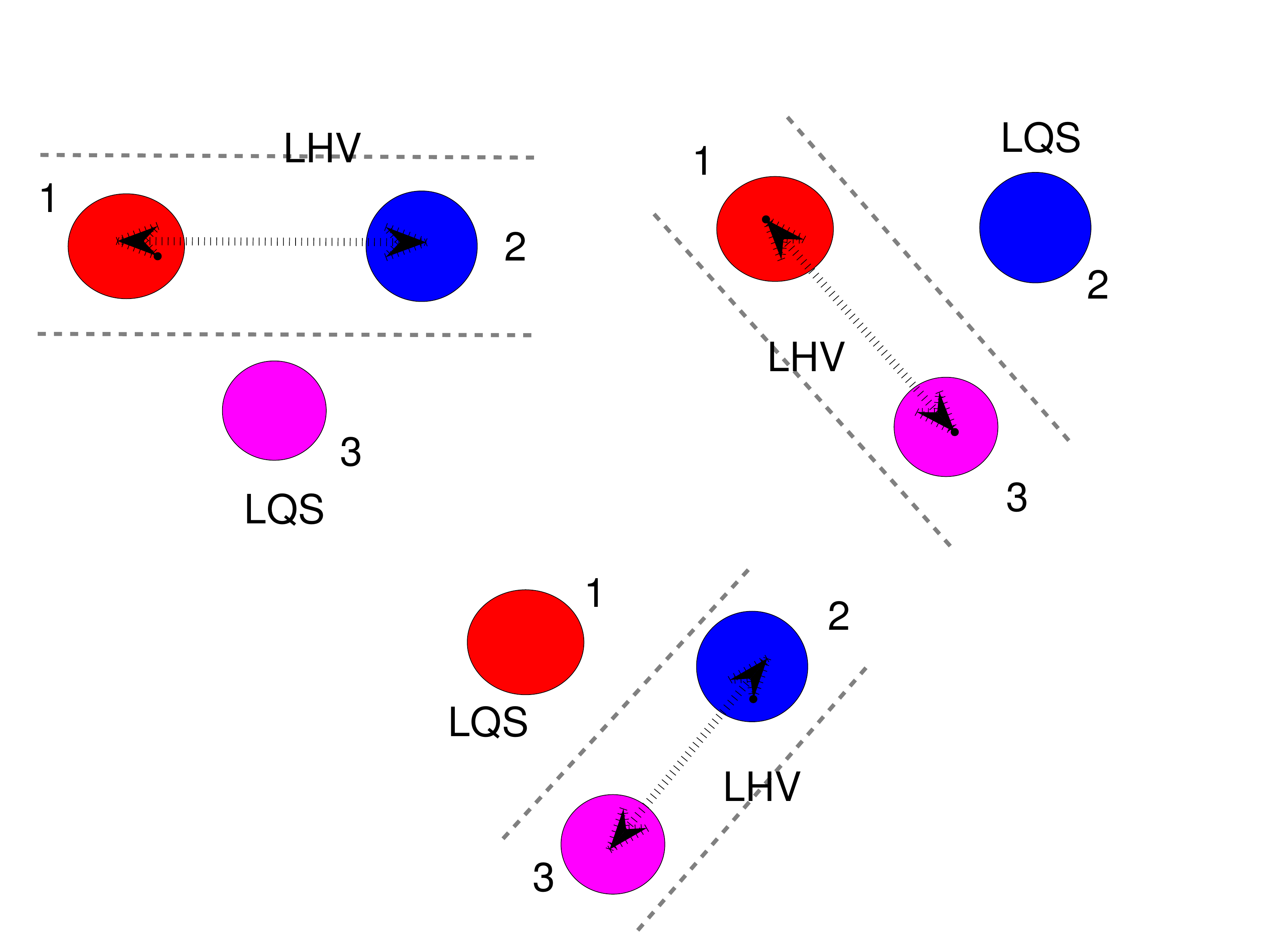}\protect\caption{(Color online) Schematic of the hybrid Local-Nonlocal Hidden States
that if negated signify genuine tripartite steering. The three depictions
are hidden variable models, in which nonlocality is allowed between
any two of the three systems (labeled $1,2,3$). The top left model
is denoted $\{12,3\}_{st}$: Here, nonlocality and therefore steering
(arrows) is allowed between systems $1,2$, which together form a
local hidden variable state (LHV). There is no steering between systems
$1,2$ and system $3$, which in the model is constrained to be a
local quantum state (LQS). The other models are $\{13,2\}_{st}$ (top
right) and $\{23,1\}_{st}$ (lower). Failure of all three models (and
any probabilistic mixtures of them) will signify genuine tripartite
steering.\label{fig:Schematic-3-1}\textcolor{red}{{} }}
\end{figure}

We introduce a notation to explain this (Fig. \ref{fig:Schematic-3-1}).
For $N=3$, there are three bipartitions of the systems: $23-1$,
$13-2$, $12-3$. The biseparable LHS description given by Eq. (\ref{eq:lhs-1})
assumes bipartitions $km-n$, but where only the system $n$ need
be a quantum system. We denote these special types of bipartition
by the notation $\{23,1\}_{st}$, $\{13,2\}_{st}$, $\{12,3\}_{st}$.
Specifically, negation of the bipartition $\{23,1\}_{st}$, implies
that we cannot write the moment $\langle X_{3}X_{2}X_{1}\rangle$
in the form $\langle X_{3}X_{2}X_{1}\rangle=\sum_{R}\eta_{R}^{^{(1)}}\langle X_{3}X_{2}\rangle\langle X_{1}\rangle_{\rho}$.
This negation implies that system $1$ is ``steerable'' by system
$\{2,3\}$ \cite{hw-1}.

The key point to the derivations of the steering criteria is that
we can only assume the quantum uncertainty relation\emph{ }for \emph{some}
of the systems. This has been explained in Ref. \cite{genepr}. First,
we assume the bipartition $\{km,n\}_{s}$ where \emph{only }system
$n$ is constrained to be a quantum state. Letting $u=h_{k}x_{k}+h_{m}x_{m}+h_{n}x_{n}$
and $v=g_{k}p_{k}+g_{m}p_{m}+g_{n}p_{n}$, it can then be shown that
 the two inequalities hold (see Appendix and Ref. \cite{genepr}):
\begin{eqnarray}
(\Delta u)^{2}+(\Delta v)^{2} & \geq & 2|h_{n}g_{n}|.\label{eq:VLFresult-4}\\
\Delta u\Delta v & \geq & |h_{n}g_{n}|\label{eq:prodst-1}
\end{eqnarray}
These relations lead to criteria for genuine tripartite steering.
 In the following, we write the ``EPR steering versions'' of the
Criteria (1-6). The proofs have been given in Ref. \cite{genepr}
or else are in the Appendix. 

To understand the significance of this sort of steering, we note that
the falsification of the biseparable state $\{km-n\}_{st}$ implies
a steering of $n$ by $km$: this means entanglement can be proved
between $n$ and the group $km$, without the assumption of good devices
for systems $km$. This type of genuine tripartite steering falsifies
any possible mixture of such bipartitions, and therefore certainly
falsifies each one of them. Therefore, the genuine tripartite steering
is certainly sufficient to imply that any two parties can ``steer''
the third. In demonstrating genuine tripartite steering, it is negated
that the steering of the three-party system can be described by consideration
of two-party steering models alone. This confirms a genuine sharing
of steering among three systems, and gives insight into a fundamental
property of quantum mechanics.

\textbf{\textcolor{black}{Criteria (3s), (4s):}}\textcolor{black}{{}
}Genuine tripartite EPR steering is observed if 
\begin{equation}
[\Delta(x_{1}-\frac{(x_{2}+x_{3})}{\sqrt{2}})]^{2}+[\Delta(p_{1}+\frac{(p_{2}+p_{3})}{\sqrt{2}})]^{2}\geq1\label{eq:eprcrit1-1}
\end{equation}
is violated (Criterion (3s)), or if 
\begin{equation}
\Delta(x_{1}-\frac{(x_{2}+x_{3})}{\sqrt{2}})\times\Delta(p_{1}+\frac{(p_{2}+p_{3})}{\sqrt{2}})\geq0.5\label{eq:crit1-1-1-1-1-1-1}
\end{equation}
is violated (Criterion (4s)). These steering inequalities are used
in Figure \ref{fig:gentri3-2-3}. The proofs have been given in Refs.
\cite{genepr} and \cite{seiji2}, and are given in our notation in
the Appendix.

\textbf{\textcolor{black}{Criteria (5s), (6s): }}The violation of
either one of the inequalities\textbf{ }
\begin{eqnarray}
(\Delta u)^{2}+(\Delta v)^{2} & \geq & 2\min\{|g_{1}h_{1}|,|g_{2}h_{2}|,|g_{3}h_{3}|\},\label{eq:eprgencrti-2-1}\\
\Delta u\Delta v & \geq & \min\{|g_{1}h_{1}|,|g_{2}h_{2}|,|g_{3}h_{3}|\},\label{eq:eprgencrti-2-1-1}
\end{eqnarray}
where $u=h_{1}x_{1}+h_{2}x_{2}+h_{3}x_{3}$, $v=g_{1}p_{1}+g_{2}p_{2}+g_{3}p_{3}$
is sufficient to confirm genuine tripartite EPR steering. 

\textbf{\textcolor{black}{Proof}}\textcolor{black}{: }Using Eq. (\ref{eq:VLFresult-4}),
we see that the bipartition $\{12,3\}_{s}$ gives the constraint $(\Delta u)^{2}+(\Delta v)^{2}\geq2|g_{3}h_{3}|$;
the bipartition $\{13,2\}_{s}$ implies $ $$(\Delta u)^{2}+(\Delta v)^{2}\geq2|g_{2}h_{2}|$;
and the bipartition $\{23,1\}_{s}$ implies $(\Delta u)^{2}+(\Delta v)^{2}\geq2|g_{1}h_{1}|$.
Thus, using Eq. (\ref{eq:convesum}), for any mixture of the bipartitions,
we can say that 
\begin{equation}
(\Delta u)^{2}+(\Delta v)^{2}\geq2\min\{|g_{1}h_{1}|,|g_{2}h_{2}|,|g_{3}h_{3}|\}.\label{eq:gencrti-1}
\end{equation}
Violation of Eq. (\ref{eq:gencrti-1}) confirms genuine tripartite
steering. The product result follows similarly, from (\ref{eq:prodst-1}).
$\square$

We can simplify these Criteria. On putting $g_{1}=h_{1}=1$ and selecting
$h_{2}=h_{3}=h$ and $g_{2}=g_{3}=g$, the right side of the inequality
becomes $2\min\{1,|gh|\}$. Now, if we take $|gh|<1$ as in Table
\ref{tableBS-1}, the inequalities take the simpler form 
\begin{equation}
(\Delta u)^{2}+(\Delta v)^{2}\geq2|gh|\label{eq:geneprcrti-1simple-3}
\end{equation}
and $\Delta u\Delta v\geq|gh|$. This inequality is used to demonstrate
genuine tripartite EPR steering, in Fig. \ref{fig:gentri3-2-3}. 

It is now possible to derive a set of three ``EPR steering inequalities''
similar to those derived by van Loock and Furusawa. This has been
explained in Ref. \cite{genepr}. The assumption that the system is
in one of the bipartitions $\{km,n\}_{st}$ will lead to a ``steering
inequality'', that if violated implies system $n$ is steerable by
the combined two systems $\{km\}$. Considering each of the three
possible bipartitions, there are three ``steering'' inequalities
identical to the van Loock and Furusawa inequalities \cite{cvsig}
but with a different right-side bound:
\begin{eqnarray}
B_{I}\equiv[\Delta(x_{1}-x_{2})]^{2}+[\Delta(p_{1}+p_{2}+g_{3}p_{3})]^{2} & \geq & 2,\nonumber \\
B_{II}\equiv[\Delta(x_{2}-x_{3})]^{2}+[\Delta(g_{1}p_{1}+p_{2}+p_{3})]^{2} & \geq & 2,\nonumber \\
B_{III}\equiv[\Delta(x_{1}-x_{3})]^{2}+[\Delta(p_{1}+g_{2}p_{2}+p_{3})]^{2} & \geq & 2,\nonumber \\
\end{eqnarray}
In fact, inequality $B_{I}\geq2$ is implied by bipartitions $\{23,1\}_{s}$
and $\{13,2\}_{s}$; inequality $B_{II}\geq2$ is implied by $\{13,2\}_{s}$
and $\{12,3\}_{s}$; and inequality $B_{III}\geq2$ is implied by
$\{12,3\}_{s}$ and $\{23,1\}_{s}$. Thus, $B_{I}<2$ signifies steering
of 1 by $\{23\}$, and also steering of $2$ by $\{13\}$, etc. The
proof of these inequalities is as for the original proof of the van
Loock-Furusawa inequalities, but assuming only the uncertainty relation
$\Delta x\Delta p\geq1$ for the steered system $n$ \cite{genepr}.
A second associated set of EPR steering inequalities involving products
can also be derived: 
\begin{eqnarray}
S_{I}\equiv\Delta(x_{1}-x_{2})\Delta(p_{1}+p_{2}+g_{3}p_{3}) & \geq & 1,\nonumber \\
S_{II}\equiv\Delta(x_{2}-x_{3})\Delta(g_{1}p_{1}+p_{2}+p_{3}) & \geq & 1,\nonumber \\
S_{III}\equiv\Delta(x_{1}-x_{3})\Delta(p_{1}+g_{2}p_{2}+p_{3}) & \geq & 1,\nonumber \\
\label{eq:threeineq-1-1-1}
\end{eqnarray}
These are the steering versions of the product inequalities Eq. (\ref{eq:threeineq-1}).
Here, inequality $S_{I}\ge1$ is implied by bipartitions $\{23,1\}_{s}$
and $\{13,2\}_{s}$; inequality $S_{II}\geq1$ is implied by $\{13,2\}_{s}$
and $\{12,3\}_{s}$; and inequality $S_{III}\geq1$ is implied by
$\{12,3\}_{s}$ and $\{23,1\}_{s}$. Thus, $S_{I}<2$ signifies steering
of 1 by $\{23\}$, and also steering of $2$ by $\{13\}$, etc. 

The inequalities lead us to the steering versions of the Criteria
(1) and (2), used in Figure \ref{fig:gentri3-1}.

\textbf{\textcolor{black}{Criterion (1s), (2s): }}We confirm genuine
tripartite steering if either the inequality 
\begin{equation}
B_{I}+B_{II}+B_{III}\geq4\label{eq:threesum-2-1}
\end{equation}
or the inequality $S_{I}+S_{II}+S_{III}\geq2$ is violated. Here,
$B_{I}\geq4$ , $B_{II}\geq4$, and $B_{III}\geq4$ are the van Loock-Furusawa
inequalities, Eq. (\ref{eq:threeineq}) and $S_{I}\geq2$ , $S_{II}\geq2$,
and $S_{III}\geq2$ are the product van Loock-Furusawa-type inequalities,
Eq. (\ref{eq:threeineq-1}). We note that each of $B_{I}$, $B_{II}$,
$B_{III}$ is a function of the variable parameters $g_{1}$, $g_{2}$,
$g_{3}$, respectively. The proof is given in the Supplemental Material
of Ref. \cite{genepr}, and is given in our own notation in the Appendix.

\begin{figure}
\includegraphics[width=1\columnwidth]{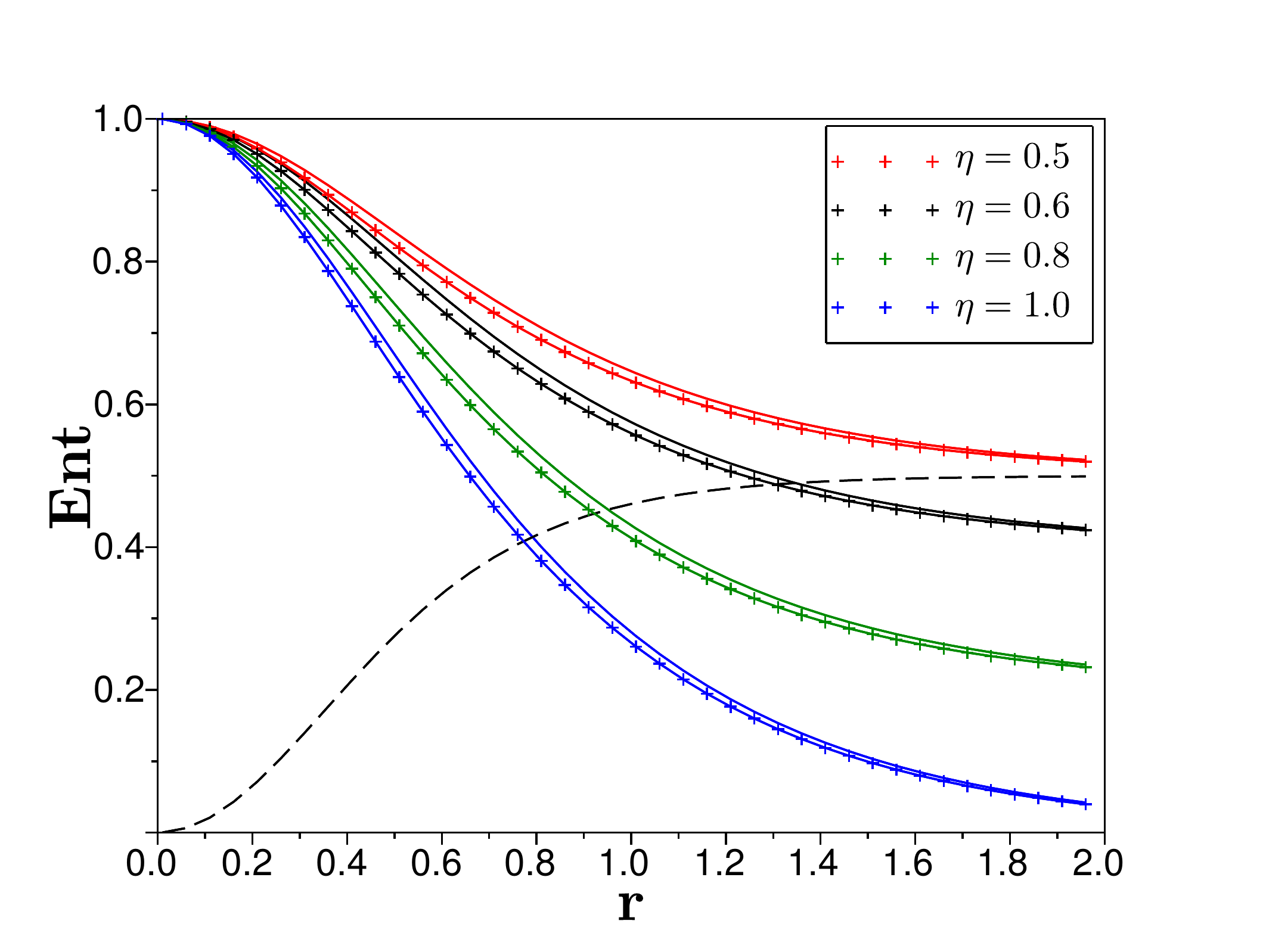}

\protect\caption{(Color online) \emph{The genuine tripartite entanglement of the CV
GHZ and EPR-type states (Figures \ref{fig:Schematic} and \ref{fig:Schematic-2})
with loss on the mode labeled $1$: }The efficiency of the beam 1
is given by $\eta$. $Ent<1$ signifies genuine tripartite entanglement.
We use the Criteria (5) and (6) with $h$ and $g$ given by Table
\ref{tableBS-1}. We signify genuine tripartite steering if $Ent$
is below the black dashed line, as given by Criteria (5s) and (6s).
The efficiency of transmission for the beam $1$ is $\eta$. The $+$
curves are for the EPR-type state using the sum or product criterion,
the two criteria giving indistinguishable results here. The second
(upper) line of each pair of the same color gives the result for the
GHZ state using the product criterion.\label{fig:gentri3}\textbf{\textcolor{red}{ }}}
\end{figure}

\begin{figure}
\includegraphics[width=1\columnwidth]{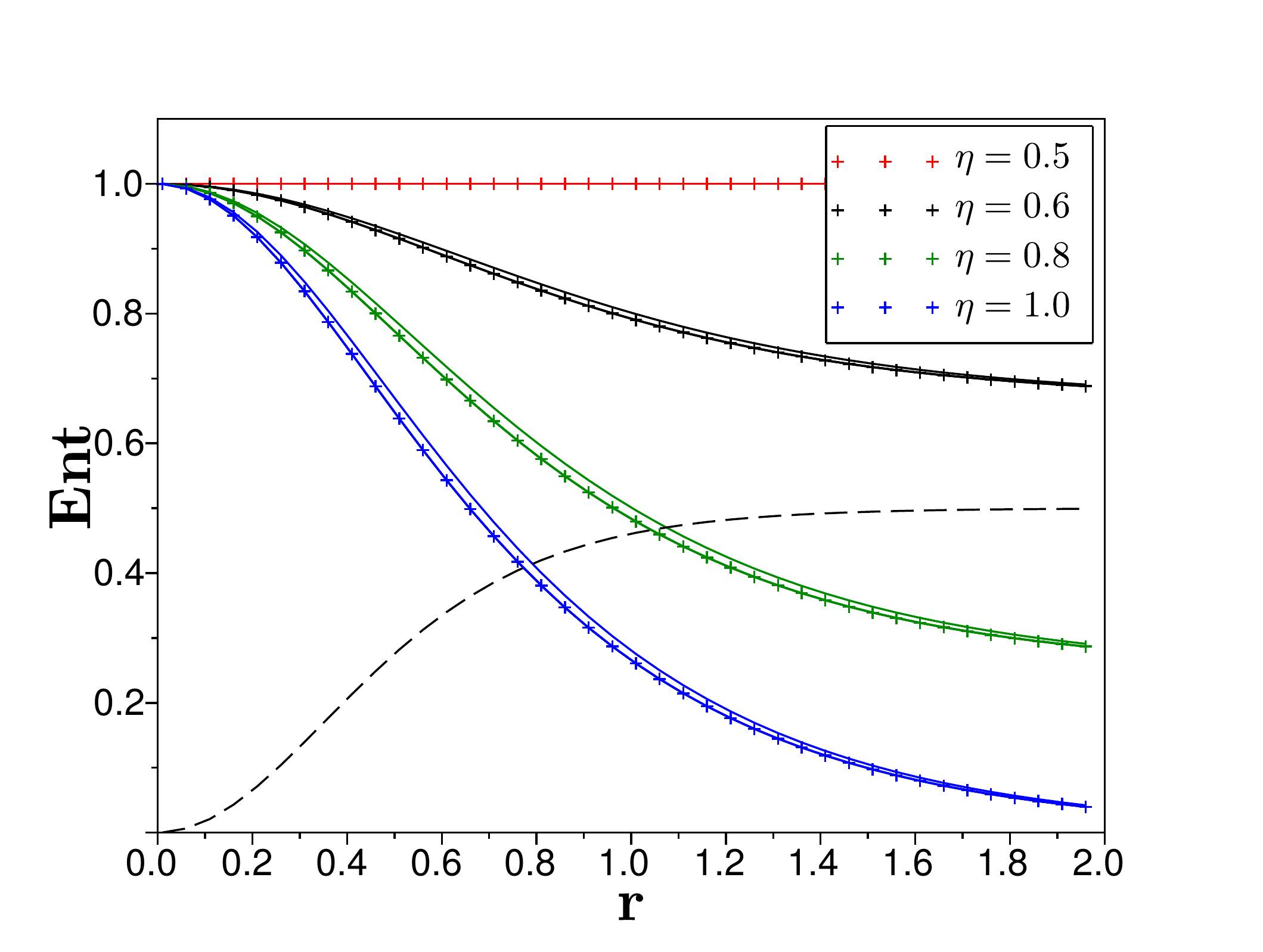}

\protect\caption{(Color online)\emph{ The genuine tripartite entanglement of the CV
GHZ and EPR-type states (Figs. \ref{fig:Schematic} and \ref{fig:Schematic-2})
with loss on the modes labeled $2$ and $3$:} $Ent<1$ signifies
genuine tripartite entanglement. Labels and curves as for Figure \ref{fig:gentri3}.
The efficiency of transmission for each of the beams 2 and 3 is given
by $\eta$.\label{fig:gentri3-4}}
\end{figure}

\section{Effect of Losses}

So far, we have only considered detection of genuine tripartite entanglement
for pure states. However these idealized states are difficult to generate
in the laboratory. There are two main sources of imperfection in the
experiments: the impurity of the input squeezed states and the losses
that occur during transmission along the channels. In this section,
we analyze the effect of losses.

The transmission losses can be modeled using a simple beam-splitter
model, in which the outputs after loss are given by $a_{out}=\sqrt{\eta}a_{in}+\sqrt{(1-\eta)}a_{vac}$,
where $a_{in}$ is the mode before loss, $a_{vac}$ is a quantum vacuum
mode, and $\eta$ is the efficiency factor that gives the altered
transmission intensity of the field mode after the loss has taken
place. 

The effect of loss on the genuine tripartite entanglement as detected
by the Criteria (5) and (6) is shown in Figs. \ref{fig:gentri3} and
\ref{fig:gentri3-4}. The most notable feature of the curves is the
loss of the criterion as $\eta\rightarrow0.5$. This can be explained
based on a knowledge of steering. Generally, we say that a system
$1$ is steerable by a group of systems labeled $B$ if we can show
$[\Delta(x_{1}-x_{B})]^{2}+[\Delta(p_{1}+p_{B})]^{2}<2$ (or $\Delta(x_{1}-x_{B})\Delta(p_{1}+p_{B})<1$)
\cite{rmp-1,hw-1,mdrepr}. Here, $x_{B}$ and $p_{B}$ can be any
measurements for system $B$. Then we note that as $r\rightarrow\infty$,
the genuine tripartite entanglement criteria used in the Figures are
given by violation of the inequalities Eqs. (\ref{eq:crit1-5}) and
(\ref{eq:crit1-2}), which are precisely of the form that signifies
steering of mode $1$ by the system $\{2,3\}$. It has also been shown
based on monogamy relations that steering cannot take place with 50\%
or more loss on the steering system (in this case, \{${2,3\}}$) \cite{monog}.
This explains the impossibility of the Criteria (5) and (6) being
satisfied (for large $r$) in Figure \ref{fig:gentri3-4} for $\eta\leq0.5$. 

We note that there is not the same restriction if we put the losses
on the steered party \cite{monog}, and hence the reduced sensitivity
to losses shown in the plots of Figure \ref{fig:gentri3}, where the
loss is entirely on party $1$. Also, we can manipulate the criteria
$Ent<1$ given by the inequalities Eq. (\ref{eq:crit1-5}) and Eq.
(\ref{eq:crit1-2}) into the form $[\Delta(x_{2}-x_{B})]^{2}+[\Delta(p_{2}+p_{B})]^{2}<4$
where now $B$ is the system containing modes $1$ and $3$. With
$50\%$ loss on the modes $1$ and $3$, we cannot demonstrate the
steering of mode $2$, which implies that $[\Delta(x_{2}-x_{B})]^{2}+[\Delta(p_{2}+p_{B})]^{2}>2$.
Thus, we will observe $Ent>0.5$ in this case. This illustrates the
asymmetry of the Criterion (5) with respect to the three parties.
In short, this means that where transmission losses on a particular
party (say $1$) are significant, it will be necessary to select the
appropriate entanglement criterion.

\section{Criteria for Genuine $N$-partite entanglement}

\begin{figure}
\includegraphics[width=0.8\columnwidth]{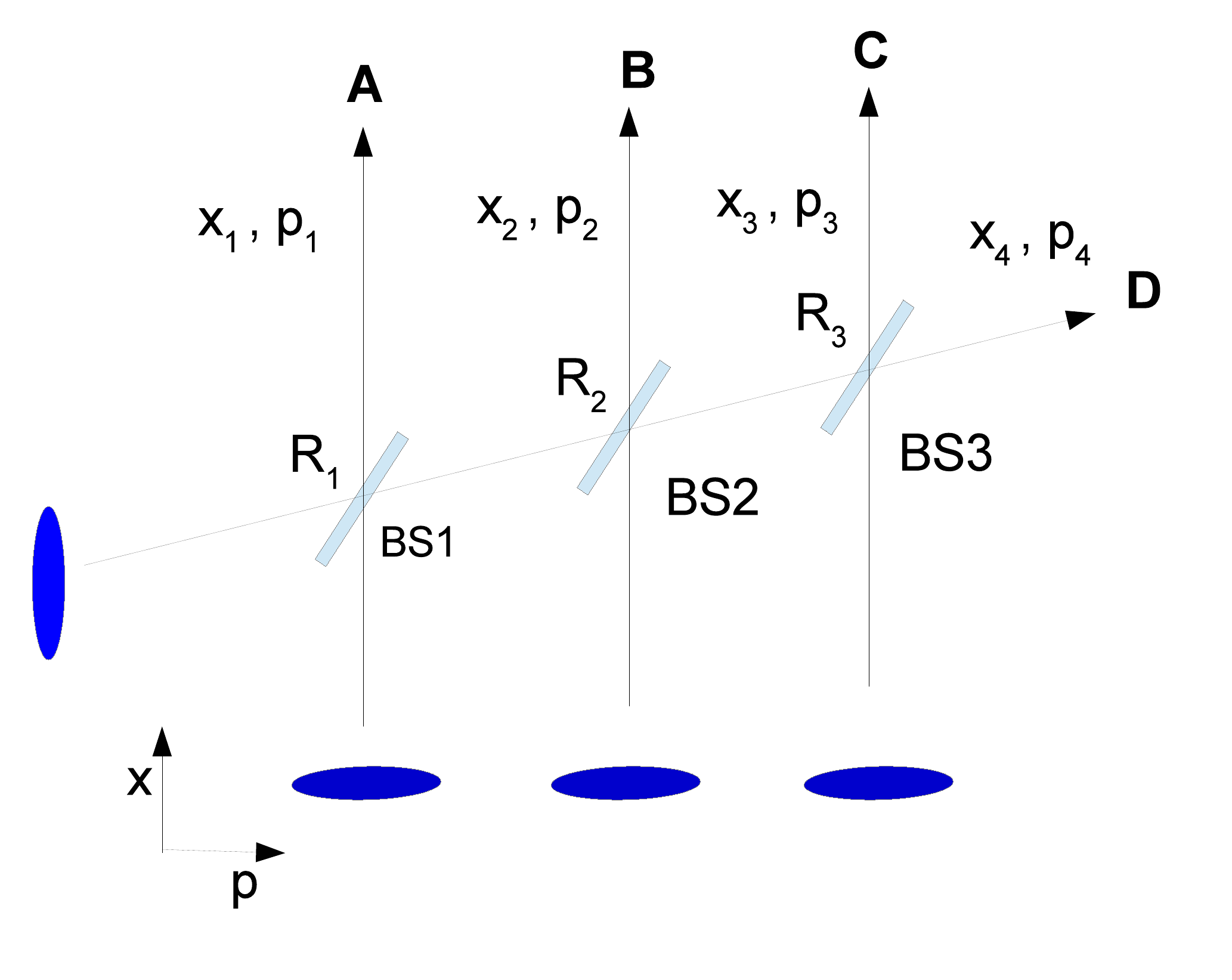}

\protect\caption{(Color online) Schematic of the generation of a genuinely four-partite
entangled state: the CV GHZ state, formed using squeezed inputs where
$R_{1}=\frac{1}{4}$, $R_{2}=\frac{1}{3}$, and $R_{3}=\frac{1}{2}$.
The generalization to arbitrary $N$ is discussed in the Ref. \cite{cvsig}.\label{fig:Schematic-1}}
\end{figure}

The above approaches can be generalized to higher $N$. Genuine $N$-partite
entangled states can be generated by extending the schemes of Fig.
\ref{fig:Schematic}, as explained in Refs. \cite{cvsig,seiji} and
depicted in Figs. \ref{fig:Schematic-1}-\ref{fig:Schematic-1-1-1}
for $N=4$. To prove genuine $N$-partite entanglement, one needs
to negate all mixtures of the biseparable states, as explained in
Sec. II. In this section, we consider three types of multipartite
entangled states, as depicted for $N=4$ in Figs. \ref{fig:Schematic-1}-\ref{fig:Schematic-1-1-1}.
The first are the \emph{CV GHZ states}, studied in Refs. \cite{cvsig,braunghz,aokicv},
and generated by successively applying beam splitters to one of the
entangled modes, with specified squeezed inputs. The second are the
\emph{asymmetric EPR-type states I}, studied in Ref. \cite{cvsig}
and formed by a sequence of beam splitters applied to \emph{one} of
the original two entangled modes. These states are depicted in Figure
\ref{fig:Schematic-1-1} for $N=4$. The third are the alternative
EPR-type states, that we call \emph{symmetric EPR-type states II},
formed by applying successive beam splitters to both arms of the entangled
pair (Figure \ref{fig:Schematic-1-1-1}). These have been generated
in Ref. \cite{seiji}. 

\subsection{Criteria for $N$-partite entanglement that use a single inequality}

First, we extend the method described in the earlier sections, and
look for a single inequality (involving just two variances) that may
be effective as criterion for detecting the genuine $N$- partite
entanglement. As we learned from the previous sections, we expect
the best choice of inequality will be related to how the entangled
state is generated. 

\begin{figure}
\includegraphics[width=0.8\columnwidth]{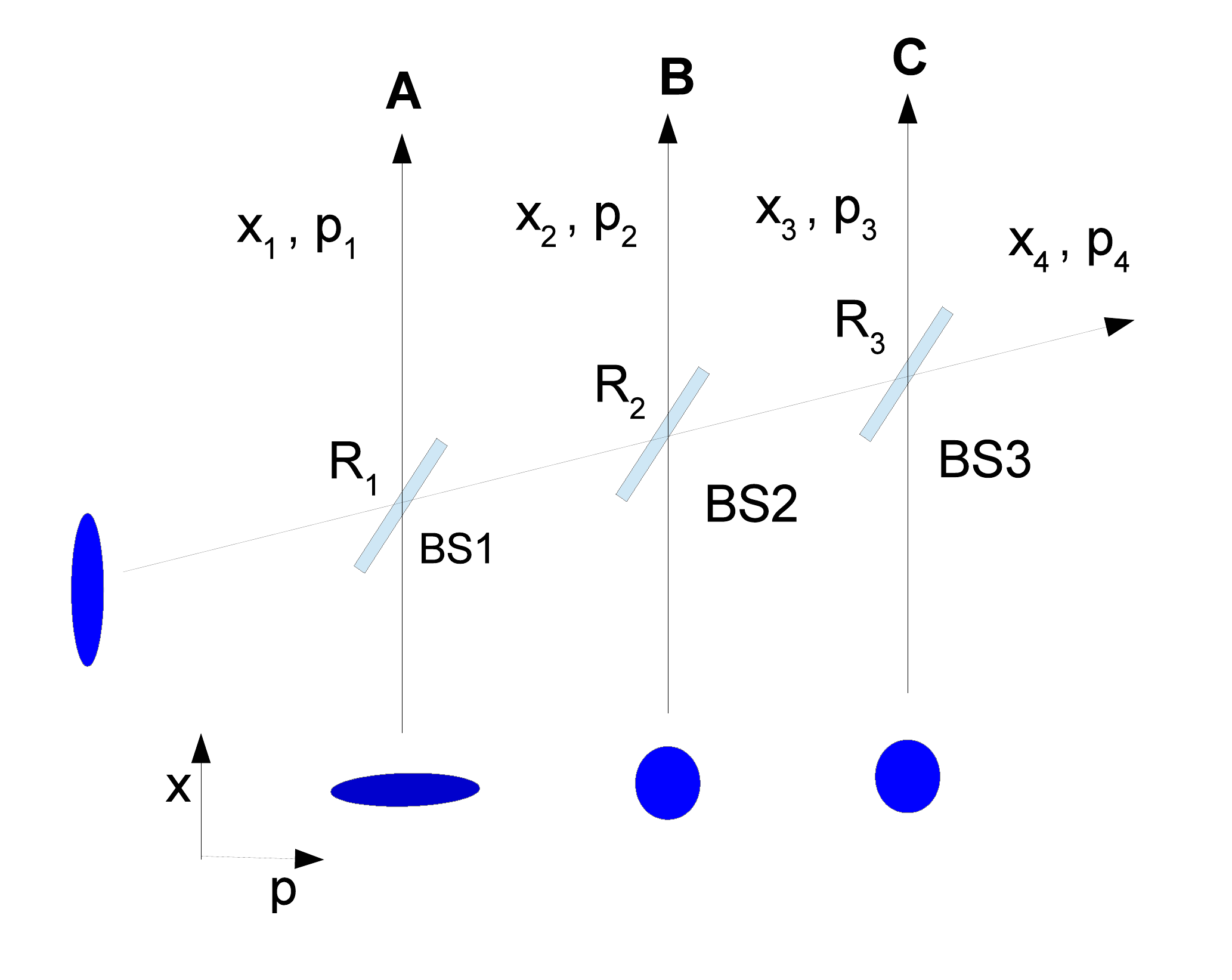}\protect\caption{(Color online) Schematic of the generation of a genuinely four-partite
entangled state: the asymmetric EPR-type state $I$ formed using a
vacuum input at all but the first beam splitter and with $R_{1}=\frac{1}{2},R_{2}=\frac{1}{3},R_{3}=\frac{1}{2}.$
By applying a further sequence of beam splitters $R_{1}=\frac{1}{2},R_{2}=\frac{1}{N-1},R_{3}=\frac{1}{(N-1)-1}...,R_{N-1}=\frac{1}{(N-1)-(N-3)}$,
as explained in the Ref. \cite{cvsig}, this state can be generalized
to arbitrary $N$.\label{fig:Schematic-1-1} }
\end{figure}

Van Loock and Furusawa \cite{cvsig} considered the following inequality
for $u=x_{1}-\frac{1}{\sqrt{N-1}}(\sum_{i=2}^{N}x_{i})$ and $v=p_{1}+\frac{1}{\sqrt{N-1}}(\sum_{i=2}^{N}p_{i})$.
They showed that 
\begin{equation}
(\Delta u)^{2}+(\Delta v)^{2}\geq\frac{4}{(N-1)}\label{eq:crit1-1}
\end{equation}
is satisfied by all biseparable states in the $N$ mode case. Hence,
using Result (2) given by Eq. (\ref{eq:single}), we deduce that violation
of this inequality will be sufficient to signify genuine $N$-partite
entanglement. This will be useful to detect the $N$-partite entanglement
of the asymmetric EPR-type state I, depicted for $N=4$ in Figure
\ref{fig:Schematic-1-1}.
 
\begin{figure}
\includegraphics[width=0.8\columnwidth]{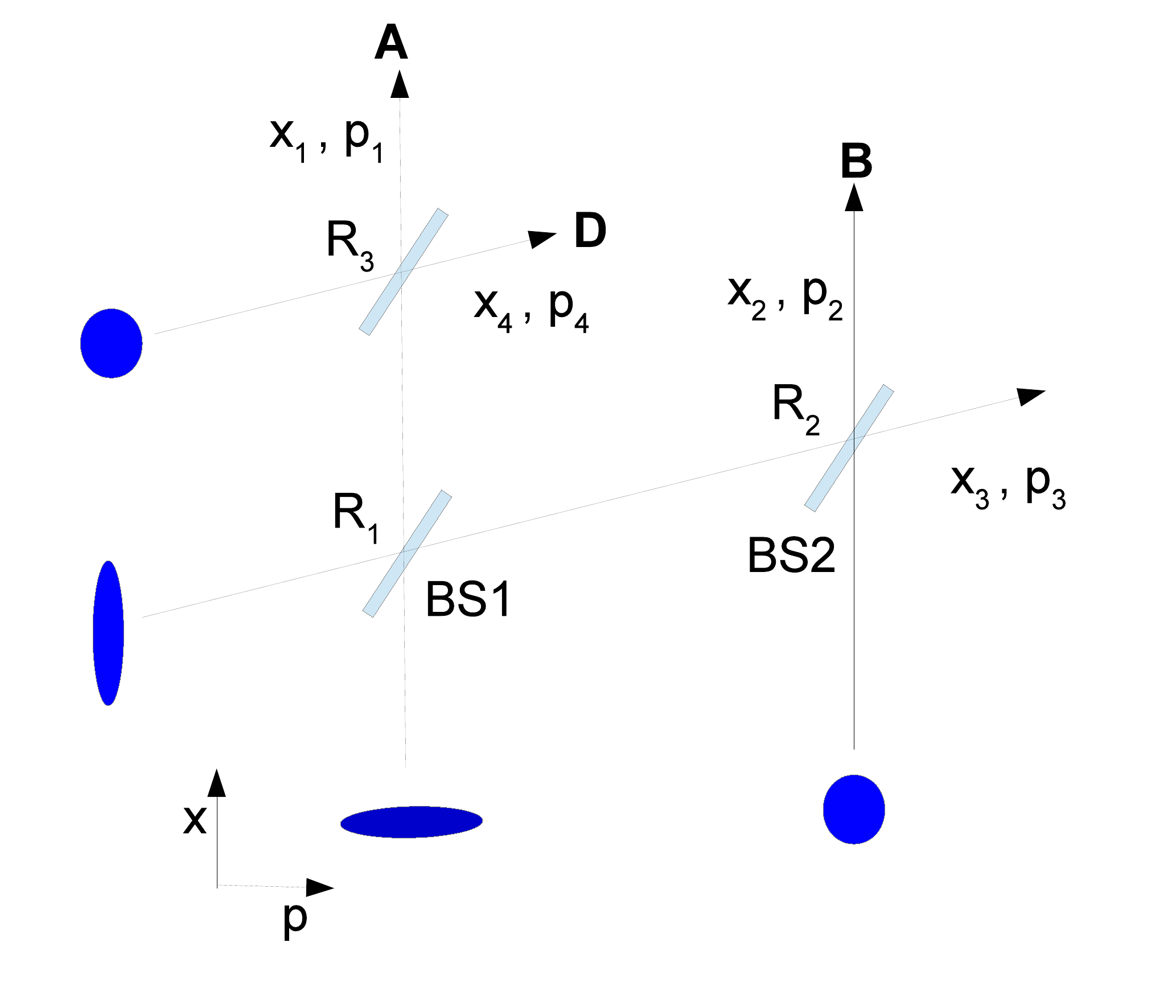}\protect\caption{(Color online) Schematic of the generation of a genuinely four-partite
entangled state: the symmetric EPR-type state $II$ created when $R_{1}=\frac{1}{2}$,
$R_{2}=\frac{1}{2}$ and $R_{3}=\frac{1}{2}$. The generalization
to arbitrary $N$ is discussed in the Ref. \cite{seiji}\label{fig:Schematic-1-1-1}}
\end{figure}

Here, we generalize the inequality (\ref{eq:crit1-1}), deriving a
criterion that is also useful to detect the multipartite entanglement
of the second type of EPR-type state II for $N=4$.

\textbf{\textcolor{black}{Criterion (8): }}We define $u=\sum_{i}h_{i}x_{i}$
and $v=\sum_{k}g_{k}p_{k}$ (although will take $h_{1}=g_{1}=1$).
For $N$ modes, suppose there are $X_{N}$ possible bipartitions.
The bipartitions in the four-mode case are $123-4$, $124-3$, $234-1$,
$134-2$, $12-34$, $13-24$, $14-23$. We can symbolize each bipartition
by $S_{r}-S_{s}$ where $S_{r}$ and $S_{s}$ are two disjoint sets
of modes so that their union is the whole set of $N$ modes. We index
the first set $S_{r}$ by $k_{r}=1,\cdots,m$ and the second set $S_{s}$
by $k_{s}=1,\cdots,n$, and we note that $n+m=N$. The violation of
the single inequality 
\begin{equation}
(\Delta u)^{2}+(\Delta v)^{2}\geq2min\{S_{B}\},\label{eq:crit1-1-4-1}
\end{equation}
where $S_{B}$ is the set of the numbers $\Bigl(\Bigl|\sum_{k_{r}=1}^{m}h_{k_{r}}g_{k_{r}}\Bigr|+\Bigl|\sum_{k_{s}=1}^{n}h_{k_{s}}g_{k_{s}}\Bigr|\Bigr)$
evaluated for each bipartition $S_{r}-S_{s}$, is sufficient to demonstrate
$N$-partite entanglement. For the figures, we define for this criterion,
$Ent=\{(\Delta u)^{2}+(\Delta v)^{2}\}/(2min\{S_{B}\})$.

\textbf{\textcolor{black}{Proof:}} Van Loock and Furusawa have shown
\cite{cvsig} that the partially separable bipartition $\rho=\sum_{R}\eta_{R}\rho_{S_{r}}^{R}\rho_{S_{s}}^{R}$
will imply
\begin{equation}
(\Delta u)^{2}+(\Delta v)^{2}\geq2\Bigl(\Bigl|\sum_{k_{r}=1}^{m}h_{k_{r}}g_{k_{r}}\Bigr|+\Bigl|\sum_{k_{s}=1}^{n}h_{k_{s}}g_{k_{s}}\Bigr|\Bigr)\label{eq:crit1-1-3}
\end{equation}
Then, we use the Result (2) (Eq. (\ref{eq:single})) and follow the
logic of the proof for Criterion (5). $\square$

Specifically, for $N=4$, we see that the inequality of Criterion
(8) reduces to:

\begin{multline}
(\Delta u)^{2}+(\Delta v)^{2}\\
\geq2min\{\Bigl|h_{1}g_{1}+h_{2}g_{2}+h_{3}g_{3}\Bigr|+\Bigl|h_{4}g_{4}\Bigr|,\qquad\qquad\qquad\qquad\\
\Bigl|h_{4}g_{4}+h_{3}g_{3}+h_{2}g_{2}\Bigr|+\Bigl|h_{1}g_{1}\Bigr|,\Bigl|h_{4}g_{4}+h_{1}g_{1}+h_{3}g_{3}\Bigr|+\Bigl|h_{2}g_{2}\Bigr|,\\
\Bigl|h_{4}g_{4}+h_{1}g_{1}+h_{2}g_{2}\Bigr|+\Bigl|h_{3}g_{3}\Bigr|,\Bigl|h_{1}g_{1}+h_{2}g_{2}\Bigr|+\Bigl|h_{3}g_{3}+h_{4}g_{4}\Bigr|,\\
\Bigl|h_{1}g_{1}+h_{3}g_{3}\Bigr|+\Bigl|h_{2}g_{2}+h_{4}g_{4}\Bigr|,\Bigl|h_{3}g_{3}+h_{2}g_{2}\Bigr|+\Bigl|h_{1}g_{1}+h_{4}g_{4}\Bigr|\}
\end{multline}

\begin{figure}
\includegraphics[width=1\columnwidth]{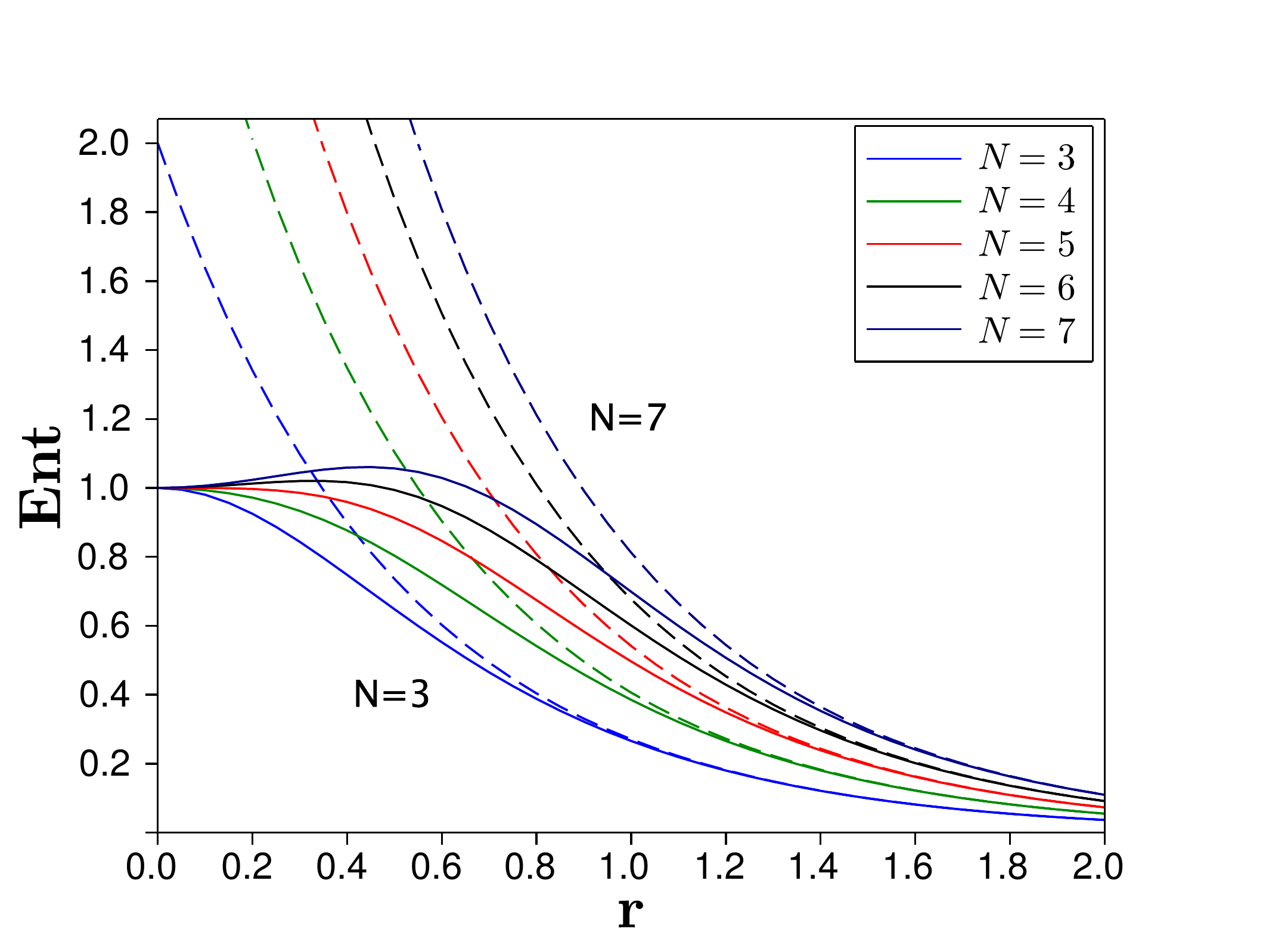}

\protect\caption{(Color online) Genuine N-partite entanglement for the asymmetric EPR-type
state I described by Fig. \ref{fig:Schematic-1-1}. Labels as in previous
figures. Here, $Ent<1$ signifies genuine $N$-partite entanglement,
using the simple criterion of Eq. (\ref{eq:crit1-1}) (dashed line)
and the generalised Criterion (8) (solid line) for $N=3-7$ (lower
to top). The values of $g_{i}$ and $h_{i}$ are given in Table \ref{tableBS-1-1-1}
.\label{fig:gentri1-2-1-1-1} }
\end{figure}

Choosing $g_{1}=h_{1}=1$, $g_{3}=-h_{3}=g_{2}=-h_{2}=g_{4}=-h_{4}=\frac{1}{\sqrt{3}}$,
we see that all biseparable states satisfy 
\begin{equation}
(\Delta u)^{2}+(\Delta v)^{2}\geq\frac{4}{3}\label{eq:n=00003D4critsinglecluster1}
\end{equation}
Violation of this inequality therefore signifies genuine $4$-partite
entanglement, which is useful for detecting the $4$-partite entanglement
of the EPR-type state I as $r\rightarrow\infty$ (Figure \ref{fig:gentri1-2-1-1-1}).

\begin{table}[h]
\protect\caption{Gains for the single inequality Criterion (8), as used for the asymmetric
EPR-type state I. \label{tableBS-1-1-1}}

\begin{tabular}{|c|c|c|c|c|c|c|}
\hline 
\multirow{2}{*}{\textbf{r}} & \multicolumn{2}{c|}{\textbf{N=4}} & \multicolumn{2}{c|}{\textbf{N=5}} & \multicolumn{2}{c|}{\textbf{N=6}}\tabularnewline
\cline{2-7} 
 & \textbf{g} & \textbf{h} & \textbf{g} & \textbf{h} & \textbf{g} & \textbf{h}\tabularnewline
\hline 
0\textbf{ } & 0 & 0 & 0 & 0 & 0 & 0\tabularnewline
\hline 
0.25\textbf{ } & 0.27 & -0.27 & 0.23 & -0.23 & 0.21 & -0.21\tabularnewline
\hline 
0.5 & 0.44 & -0.44 & 0.38 & -0.38 & 0.34 & -0.34\tabularnewline
\hline 
0.75 & 0.52 & -0.52 & 0.45 & -0.45 & 0.40 & -0.40\tabularnewline
\hline 
1 & 0.56 & -0.56 & 0.48 & -0.48 & 0.43 & -0.43\tabularnewline
\hline 
1.5 & 0.57 & -0.57 & 0.50 & -0.50 & 0.45 & -0.45\tabularnewline
\hline 
2\textbf{ } & 0.58 & -0.58 & 0.50 & -0.50 & 0.45 & -0.45\tabularnewline
\hline 
\end{tabular}
\end{table}

\begin{figure}
\includegraphics[width=1\columnwidth]{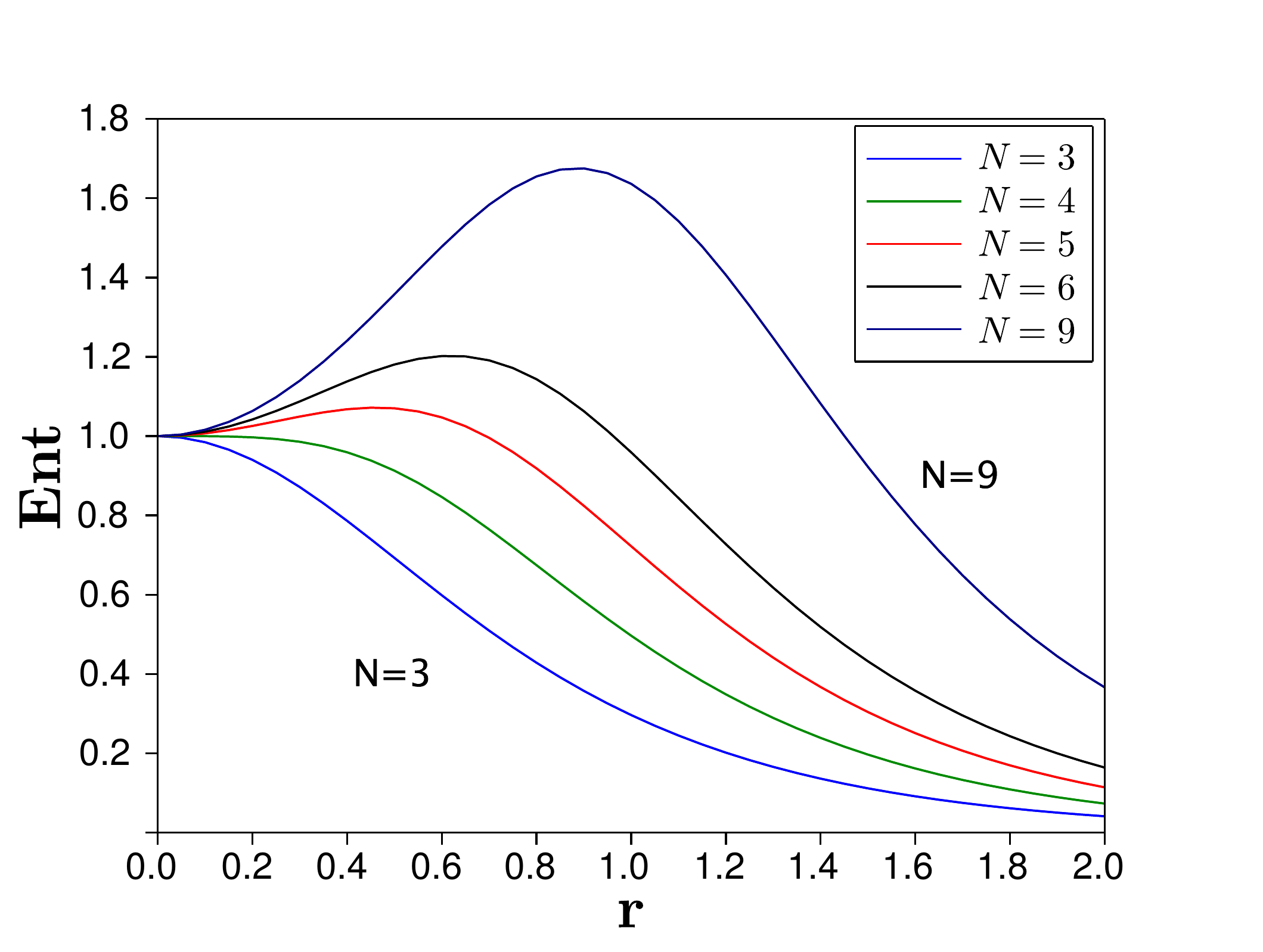}

\protect\caption{(Color online) Genuine $N$-partite entanglement using the Criterion
(8) for the CV GHZ state. Labels as in previous figures. The values
of $g_{i}$ and $h_{i}$ are given in Table \ref{tableBS-1-1}.\label{fig:gentri1-2-1-1}
\textcolor{red}{}}
\end{figure}

\begin{table}[h]
\protect\caption{Gains for single inequality Criterion (8) for the CV GHZ. Here, $h_{1}=g_{1}=1,h_{2}=h_{3}=h_{4}=h,g_{2}=g_{3}=g_{4}=g$.\label{tableBS-1-1}}

\begin{tabular}{|c|>{\centering}p{0.1\columnwidth}|>{\centering}p{0.1\columnwidth}|>{\centering}p{0.1\columnwidth}|>{\centering}p{0.1\columnwidth}|>{\centering}p{0.1\columnwidth}|>{\centering}p{0.1\columnwidth}|}
\hline 
\multirow{2}{*}{\textbf{r}} & \multicolumn{2}{c|}{\textbf{N=4}} & \multicolumn{2}{c|}{\textbf{N=5}} & \multicolumn{2}{c|}{\textbf{N=6}}\tabularnewline
\cline{2-7} 
 & \textbf{g} & \textbf{h} & \textbf{g} & \textbf{h} & \textbf{g} & \textbf{h}\tabularnewline
\hline 
0\textbf{ } & 0 & 0 & 0 & 0 & 0 & 0\tabularnewline
\hline 
0.25\textbf{ } & 0.30 & -0.19 & 0.26 & -0.14 & 0.22 & -0.12\tabularnewline
\hline 
0.5 & 0.61 & -0.28 & 0.56 & -0.21 & 0.52 & -0.17\tabularnewline
\hline 
0.75 & 0.83 & -0.31 & 0.79 & -0.23 & 0.76 & -0.19\tabularnewline
\hline 
1 & 0.93 & -0.33 & 0.91 & -0.24 & 0.90 & -0.20\tabularnewline
\hline 
1.5 & 0.99 & -0.33 & 0.99 & -0.25 & 0.99 & -0.20\tabularnewline
\hline 
2\textbf{ } & 1.00 & -0.33 & 1.00 & -0.25 & 1.00 & -0.20\tabularnewline
\hline 
\end{tabular}
\end{table}

\begin{figure}
\includegraphics[width=1\columnwidth]{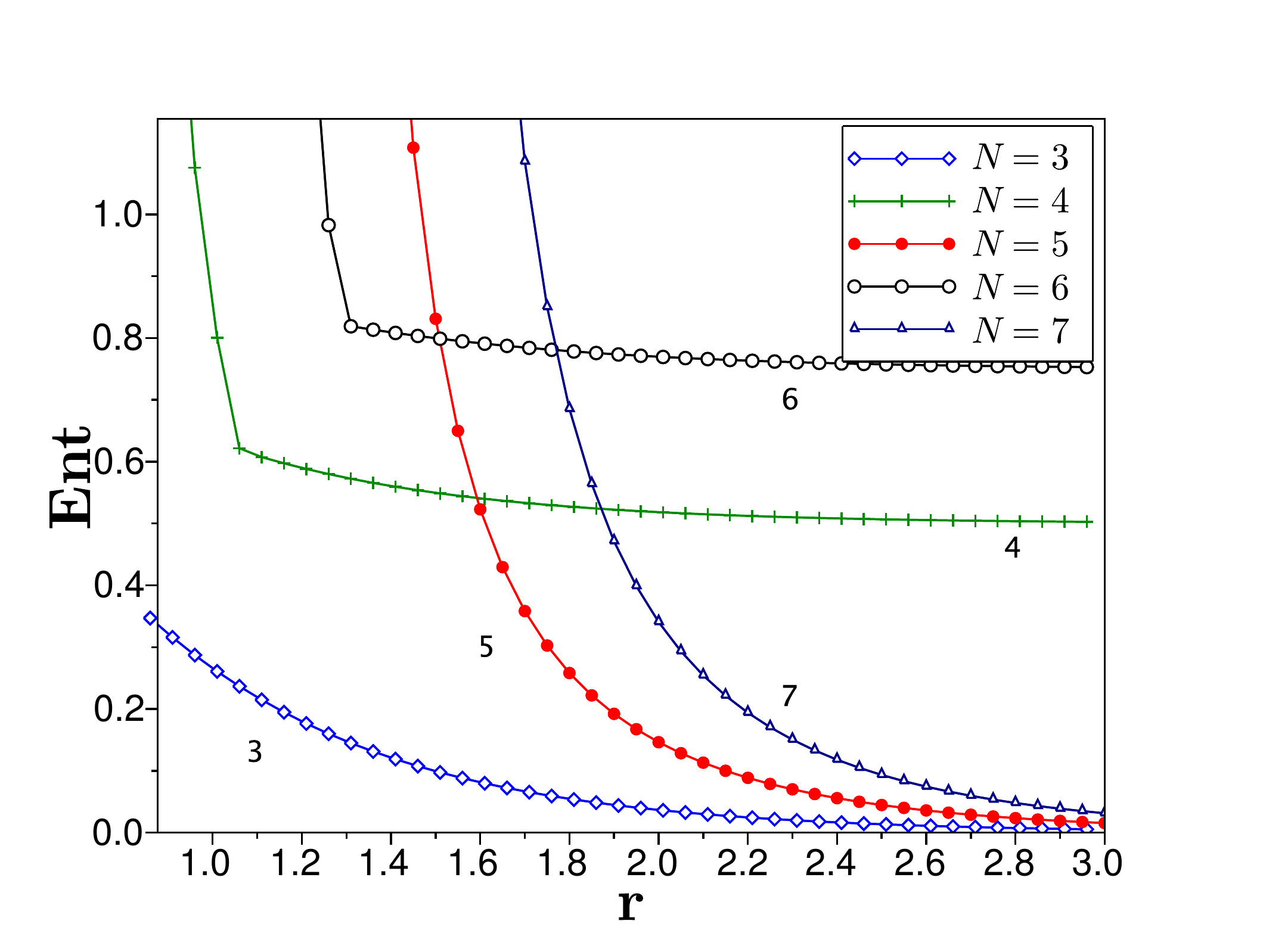}

\protect\caption{(Color online) Genuine $N$-partite entanglement for the CV symmetric
EPR-type state II of Fig. \ref{fig:Schematic-1-1-1}. Labels as in
previous figures. Green solid line corresponds to the Criterion (10)
with $N=4$; diamonds correspond to the single inequality Criteria
(8), with the values $g_{i}$, $h_{i}$ given in Table \ref{tableBS-1-1-2}.
\label{fig:gentri1-2-2}}
\end{figure}

We evaluate in Figs. \ref{fig:gentri1-2-1-1-1} - \ref{fig:gentri1-2-2}
the results of the Criteron (8) for the states generated by the networks
of Figs. \ref{fig:Schematic-1}-\ref{fig:Schematic-1-1-1}. For the
asymmetric EPR-type state I (Fig. \ref{fig:Schematic-1-1}), the simple
criterion Eq. (\ref{eq:crit1-1}) suffices to detect the $N$-partite
entanglement, as $r\rightarrow\infty$.\textcolor{red}{{} }The correlations
of this state are such that the result of $x_{1}$ (or $p_{1}$) can
be inferred from the measurement of the linear combination of the
$x_{i}$ (or $p_{i}$) of the modes on the other side of the first
beam splitter $BS1$. This leads to ideal EPR-type correlations where
both the variances of the inequality (\ref{eq:crit1-1}) go to $0$
(as $r\rightarrow\infty$) and the simple inequality is violated.
The inequality works for larger $N$, for the states generated with
specific choices of reflectivities for the beam splitter sequences
as given in Refs. \cite{seiji2,seiji}. Further, the optimization
for small $r$ is possible. The details are given in the Appendix.
\begin{table}[h]
\protect\caption{Gains for single inequality Criterion (8) for the symmetric EPR-type
II state. Here $h_{1}=g_{1}=1,h_{2}=h_{3}=...=h_{R},g_{2}=g_{3}=...g_{R},h_{4}=h_{6}=...=h_{L}=g_{4}=g_{6}=...=g_{L}$.\label{tableBS-1-1-2}}

\begin{tabular}{|>{\centering}p{0.06\columnwidth}|c|c|c|c|c|c|c|c|c|}
\hline 
\multirow{2}{0.06\columnwidth}{\textbf{r}} & \multicolumn{3}{c|}{\textbf{N=4}} & \multicolumn{3}{c|}{\textbf{N=5}} & \multicolumn{3}{c|}{\textbf{N=6}}\tabularnewline
\cline{2-10} 
 & $h_{R}$ & $h_{L}$ & $g_{R}$ & $h_{R}$ & $h_{L}$ & $g_{R}$ & $h_{R}$ & $h_{L}$ & $g_{R}$\tabularnewline
\hline 
0\textbf{ } & 0 & 0 & 0 & 0 & 0 & 0 & 0 & 0 & 0\tabularnewline
\hline 
0.25\textbf{ } & -0.24 & -0.06 & 0.24 & -0.20 & -0.06 & 0.20 & -0.17 & -0.04 & 0.17\tabularnewline
\hline 
0.5 & -0.46 & -0.21 & 0.46 & -0.38 & -0.21 & 0.38 & -0.33 & -0.15 & 0.33\tabularnewline
\hline 
0.75 & -0.63 & -0.40 & 0.63 & -0.52 & -0.40 & 0.52 & -0.50 & -0.31 & 0.50\tabularnewline
\hline 
1 & -0.76 & -0.58 & 0.76 & -0.62 & -0.58 & 0.62 & -0.63 & -0.50 & 0.63\tabularnewline
\hline 
1.5 & -0.91 & -0.82 & 0.91 & -0.74 & -0.82 & 0.74 & -0.83 & -0.75 & 0.83\tabularnewline
\hline 
2\textbf{ } & -0.96 & -0.93 & 0.96 & -0.79 & -0.93 & 0.79 & -0.93 & -0.90 & 0.93\tabularnewline
\hline 
\end{tabular}
\end{table}

The CV GHZ state (Figure \ref{fig:Schematic-1}) can also be detected
using the single inequality of Criterion (8), provided the coefficients
$g_{i}$ and $h_{i}$ are selected appropriately, as in Table \ref{tableBS-1-1}.
This choice can be determined from substitution and differentiation
to minimise the left side of the inequality. In this case, the right
side of the inequality reduces to $2[1+(N-3)gh]$. The details are
given in the Appendix, and results are presented in Figure \ref{fig:gentri1-2-1-1}.

For the symmetric EPR state II (Figure \ref{fig:Schematic-1-1-1}),
it is not as easy to find a simple single inequality that will signify
four-partite entanglement, over the entire range of $r$. The problem
is as follows: For large $r$, on examining the generation scheme
and defining the modes as in Section V.A, we note: for $BS2$, $a_{2'}=\frac{1}{\sqrt{2}}(a_{2}+a_{3})$;
for the third $BS$, $a_{1'}=\frac{1}{\sqrt{2}}(a_{1}-a_{4})$ and
hence $x'_{1}=\frac{1}{\sqrt{2}}(x_{1}-x_{4})$ and $p'_{1}=\frac{1}{\sqrt{2}}(p_{1}-p_{4})$.
This means that the original EPR correlation corresponding to $[\Delta(x'_{1}-x'_{2})]^{2}\rightarrow0$,
$[\Delta(p'_{1}+p'_{2})]^{2}\rightarrow0$, becomes $[\Delta((x_{1}-x_{4})-(x_{2}+x_{3}))]^{2}\rightarrow0$,
$[\Delta(p_{1}-p_{4}+p_{2}+p_{3})]^{2}\rightarrow0$. Thus, we can
see that these latter two variances will vanish, implying violation
of the inequality 
\begin{eqnarray}
I & \geq & 4\label{eq:4-partite some bipartitions}
\end{eqnarray}
where $I=[\Delta((x_{1}-x_{4})-(x_{2}+x_{3}))]^{2}+[\Delta(p_{1}-p_{4}+p_{2}+p_{3})]^{2}$.
Now, we see that this violation will negate biseparability of the
state with respect to the bipartitions $123-4$, $124-3$, $431-2$,
$ $$ $$234-1$, $14-23$ (use the proof of Criterion (8)). However,
the violation \emph{cannot }negate the bipartitions $12-34$ and $13-24$,
and cannot therefore demonstrate genuine four-partite entanglement.
Despite that, our analysis with general coefficients using Criterion
(8) reveals that all the bipartitions \emph{can} be negated, for a
different choice of coefficients $g_{i}$ and $h_{i}$, provided $r\rightarrow\infty$.
This means we can use the single inequality to detect genuine $N$-partite
entanglement, for highly squeezed inputs, as shown in Fig. \ref{fig:gentri1-2-2}.

\subsection{Criteria for four-partite entanglement using the van Loock-Furusawa
inequalities}

We can apply the approach of Result (1) (Eq. (\ref{eq:single})) and
Criterion (1) to derive a criterion for genuine four-partite entanglement,
based on summation of van Loock-Furusawa inequalities. We will consider
four systems, and label the set of bipartitions $123-4$, $124-3$,
$234-1$, $134-2$, $12-34$, $13-24$, $14-23$ by $k=1,..,7$. Van
Loock and Furusawa derived a set of six inequalities \cite{cvsig},
that if violated eliminate biseparability with respect to certain
bipartitions: 
\begin{eqnarray}
B_{I}\equiv[\Delta(x_{1}-x_{2})]^{2}\,\,\,\,\,\,\,\,\,\,\,\,\,\,\,\,\,\,\,\,\,\,\,\,\,\,\,\,\,\,\,\,\,\,\,\,\,\,\,\,\,\,\,\,\,\,\,\,\,\,\,\,\nonumber \\
{}+[\Delta(p_{1}+p_{2}+g_{3}p_{3}+g_{4}p_{4})]^{2} & \geq & 4\nonumber \\
B_{II}\equiv[\Delta(x_{2}-x_{3})]^{2}\,\,\,\,\,\,\,\,\,\,\,\,\,\,\,\,\,\,\,\,\,\,\,\,\,\,\,\,\,\,\,\,\,\,\,\,\,\,\,\,\,\,\,\,\,\,\,\,\,\,\,\,\nonumber \\
{}+[\Delta(g_{1}p_{1}+p_{2}+p_{3}+g_{4}p_{4})]^{2} & \geq & 4\nonumber \\
B_{III}\equiv[\Delta(x_{1}-x_{3})]^{2}\,\,\,\,\,\,\,\,\,\,\,\,\,\,\,\,\,\,\,\,\,\,\,\,\,\,\,\,\,\,\,\,\,\,\,\,\,\,\,\,\,\,\,\,\,\,\,\,\,\,\,\,\nonumber \\
{}+[\Delta(p_{1}+g_{2}p_{2}+p_{3}+g_{4}p_{4})]^{2} & \geq & 4\nonumber \\
B_{IV}\equiv[\Delta(x_{3}-x_{4})]^{2}\,\,\,\,\,\,\,\,\,\,\,\,\,\,\,\,\,\,\,\,\,\,\,\,\,\,\,\,\,\,\,\,\,\,\,\,\,\,\,\,\,\,\,\,\,\,\,\,\,\,\,\,\nonumber \\
{}+[\Delta(g_{1}p_{1}+g_{2}p_{2}+p_{3}+p_{4})]^{2} & \geq & 4\nonumber \\
B_{V}\equiv[\Delta(x_{2}-x_{4})]^{2}\,\,\,\,\,\,\,\,\,\,\,\,\,\,\,\,\,\,\,\,\,\,\,\,\,\,\,\,\,\,\,\,\,\,\,\,\,\,\,\,\,\,\,\,\,\,\,\,\,\,\,\,\nonumber \\
{}+[\Delta(g_{1}p_{1}+p_{2}+g_{3}p_{3}+p_{4})]^{2} & \geq & 4\nonumber \\
B_{VI}\equiv[\Delta(x_{1}-x_{4})]^{2}\,\,\,\,\,\,\,\,\,\,\,\,\,\,\,\,\,\,\,\,\,\,\,\,\,\,\,\,\,\,\,\,\,\,\,\,\,\,\,\,\,\,\,\,\,\,\,\,\,\,\,\,\nonumber \\
{}+[\Delta(p_{1}+g_{2}p_{2}+g_{3}p_{3}+p_{4})]^{2} & \geq & 4\nonumber \\
\label{eq:n=00003D4row}
\end{eqnarray}
where $g_{i}$ is an arbitrary real number. Van Loock and Furusawa
showed that violation of any three of these inequalities will negate
that the system can be in one of the possible biseparable states,
that we denote by $\rho_{k}$ . The violation of any three inequalities
will thus signify full four-partite inseparability. A similar set
of inequalities is derived for the case of arbitrary $N$. 

As we have seen, this is not enough to negate that the system could
be in a mixture of the biseparable states $\rho_{k}$. However, we
can extend the proof of Criterion (1), to show that sufficiently strong
violations of the inequalities (as is predicted by CV GHZ states)
will confirm genuine $4$-partite entanglement.

\textbf{\textcolor{black}{Criterion (9):}}\textcolor{black}{{} }Four
systems are genuinely four-partite entangled if the inequality 
\begin{equation}
\sum_{J=1}^{6}B_{J}\geq12\label{eq:four}
\end{equation}
 is violated, where $B_{J}\geq4$, $J=I,II,...,VI$, are the van Loock-Furusawa
inequalities (\ref{eq:n=00003D4row}).For the figures, we define
for this criterion, $Ent=(\sum_{J=1}^{6}B_{J})/12$.

\textbf{Proof: }As for Criterion (1), we begin by assuming a mixture
$\rho_{BS}=\sum_{k}P_{k}\rho_{k}$ where $\rho_{k}$ is a density
operator with the bipartition indexed by $k=1,2,....7$. Van Loock
and Furusawa showed that four of the biseparable states $\rho_{k}$
predict any particular one of the inequalities, because four of the
biseparable states $\rho_{k}$ have separability with respect to the
two systems specified by the subscripts of the positions $x$ measured
in the inequality. We can write
\[
B_{I}\geq\sum_{k=1}^{7}P_{k}B_{I,k}\geq4(P_{3}+P_{4}+P_{6}+P_{7})
\]
and similarly $B_{II}\geq4(P_{2}+P_{4}+P_{5}+P_{6})$, $B_{III}\geq4(P_{2}+P_{3}+P_{5}+P_{7})$,
$B_{IV}\geq4(P_{1}+P_{2}+P_{6}+P_{7})$, $B_{V}\geq4(P_{1}+P_{4}+P_{5}+P_{7})$$ $,
$B_{VI}\geq4(P_{1}+P_{3}+P_{5}+P_{6})$. We see that $\sum_{J}B_{J}\geq12(P_{1}+P_{2}+P_{3}+P_{4}+P_{5}+P_{6}+P_{7})$,
which gives the result. $\square$ 

For symmetric systems where the $B_{J}$ are equal, we will require
$B_{J}<2$ ($50$\% reduction of the vacuum noise level) in order
to achieve Criterion (9). Predictions are given in Figure \ref{fig:gentri3-1},
for the CV GHZ state generated by the scheme of Figure \ref{fig:Schematic-1}.
A very high degree of entanglement is possible as $r\rightarrow\infty$.
The genuine $4$-partite entanglement of the CV GHZ state is detectable
using the Criterion (9), for moderate values of $r$, though greater
squeezing is required than for the $N=3$ case. The method can be
extended to higher $N$, once the van Loock-Furusawa inequalities
are known. We note the genuine $4$-partite entanglement of the EPR-type
states is not effectively detected by this criterion.

\subsection{Criteria for $4$-partite entanglement using summation of inequalities}

Let us return to the symmetric EPR-type state II, of Figure \ref{fig:Schematic-1-1-1}.
We now use the approach of Result (1) and Criterion (1) to tailor
a criterion for this state, using the van Loock-Furusawa inequalities.
For $N=4$, we have seen that the inequality (\ref{eq:4-partite some bipartitions})
given by $I\geq4$ will negate bipartitions $123-4$, $124-3$, $431-2$,
$ $$ $$234-1$, $14-23$ but \emph{not} the bipartitions $12-34$
and $13-24$. On the other hand, the van Loock-Furusawa inequality
$B_{II}\geq4$ will negate the bipartitions $12-34$, $13-24$, $124-3$,
$431-2$. It has been shown in Ref. \cite{seiji} that the EPR-type
state II does violate the van Loock-Furusawa inequality, by a small
amount. We can prove the following:

\textbf{Criterion (10):} The violation of the inequality 
\begin{equation}
I+B_{II}\geq4\label{eq:twoineseiji}
\end{equation}
is sufficient to prove genuine $4$-partite entanglement. For the
figures, we define for this criterion, $Ent=(I+B_{II})/4$.

\textbf{Proof:} If we assume a mixture $\rho_{BS}=\sum_{k}P_{k}\rho_{k}$
where $\rho_{k}$ is a density operator biseparable across the bipartition
indexed by $k=1,2,....7,$ then $I\geq4(P_{1}+P_{2}+P_{3}+P_{4}+P_{5}$)
whereas $B_{II}\geq4(P_{2}+P_{3}+P_{6}+P_{7})$. Hence for any biseparable
state the inequality will hold. $\square$ 

The combined inequality (\ref{eq:twoineseiji}) can indeed be used
to detect the genuine $4$-partite entanglement of the EPR-type state
II, and the predictions are given in the Figure \ref{fig:gentri1-2-2}.

\section{Conclusion }

This paper examines how to confirm genuine multi-partite entanglement
using continuous variable (that is, quadrature phase amplitude) measurements,
pointing out that the approach pioneered by van Loock and Furusawa
is not in itself sufficient in realistic situations, where one needs
to exclude all mixed state models. The criteria are based on the scaled
position and momentum observables of the quantized harmonic oscillator,
and thus could also be used to detect the position and momentum entanglement
associated with quantum mechanical oscillators, as done for bipartite
entanglement in the recent experiment of Ref. \cite{optomech}.

We have presented a general strategy for deriving criteria to detect
genuine $N$-partite entanglement. Further, we present specific criteria
and algorithms for the detection of the genuine $N$-partite entanglement
of CV GHZ and EPR-type states that have been realized (or proposed)
experimentally. In the GHZ case, we show that genuine tripartite entanglement
could be confirmed for noise reductions at $2/3$ the level necessary
to violate the standard van Loock-Furusawa inequalities. We also present
specific predictions for higher $N$, and consider the effect of transmission
losses which could be important to quantum communication applications.
A more significant limitation in terms of detecting the genuine multipartite
entanglement in a laboratory is likely to be the degree of impurity
of the initial squeezed inputs. This effect has not been addressed
in this paper, but has been studied in part in Ref. \cite{seiji2}. 

For three parties, we also present criteria for genuine tripartite
steering. This corresponds to a type of entanglement giving a multipartite
EPR paradox. In that case, any single party can be ``steered'' by
the other two, which means that entanglement can be confirmed between
the two groups, even when the group of two parties (or their devices)
cannot be trusted to perform proper quantum measurements. This form
of entanglement is likely to be useful to multiparty one-sided device-independent
quantum cryptography.
\begin{acknowledgments}
This work was suppported by the Australian Research Council Discovery
Projects program. We are grateful to P. Drummond, Q. He, S. Armstrong
and P.K. Lam for stimulating discussions.
\end{acknowledgments}

\section*{Appendix}

\subsection{Proof of the relation (\ref{eq:uvineq-1}) }

Let us assume that the system is described by the mixture $\rho_{km,n}=\sum_{i}\eta_{i}^{(n)}\rho_{km}^{i}\rho_{n}^{i}$.
Then on using the Cauchy Schwarz inequality, we find 
\begin{eqnarray}
(\Delta u)^{2}(\Delta v)^{2} & \geq & [\sum_{i}\eta_{i}^{(n)}(\Delta u)_{i}^{2}][\sum_{i}\eta_{i}^{(n)}(\Delta v)_{i}^{2}]\nonumber \\
 & \geq & [\sum_{i}\eta_{i}^{(n)}(\Delta u)_{i}(\Delta v)_{i}]^{2},\label{eq:csprod-1}
\end{eqnarray}
where $(\Delta u)_{i}(\Delta v)_{i}$ is the product of the variances
for a pure product state of type $\psi_{km}\psi_{n}$ denoted by $i$.
Generally, let us consider a system in a product state of type $\psi_{a}\psi_{b}$
and define the linear combinations $x_{a}+gx_{b}$ and $p_{a}+gp_{b}$
of the operators $x_{a}$, $p_{a}$ and $x_{b}$, $p_{b}$ for the
systems described by wavefunctions $\psi_{a}$ and $\psi_{b}$ respectively.
It is always true that the variances for such a product state satisfy
$[\Delta(x_{a}+gx_{b})]^{2}=(\Delta x_{a})^{2}+g^{2}(\Delta x_{b})^{2}$
and $[\Delta(p_{a}+gp_{b})]^{2}=(\Delta p_{a})^{2}+g^{2}(\Delta p_{b})^{2}$.
This implies that
\begin{eqnarray}
[\Delta(x_{a}+gx_{b})]^{2}[\Delta(p_{a}+gp_{b})]^{2} & = & [(\Delta x_{a})^{2}+g^{2}(\Delta x_{b})^{2}]\nonumber \\
 &  & \times[(\Delta p_{a})^{2}+g^{2}(\Delta p_{b})^{2}]\nonumber \\
 & \geq & [\Delta x_{a}\Delta p_{a}+g^{2}\Delta x_{b}\Delta p_{b}]^{2}\nonumber \\
\end{eqnarray}
where we use that for any real numbers $x$ and $y$, $x^{2}+y^{2}\geq2xy$.
We can apply this result to deduce that for a product state of type
$\psi_{km}\psi_{n}$, it is true that $(\Delta u)_{i}(\Delta v)_{i}\geq[\Delta(h_{k}x_{k}+h_{m}x_{m})][\Delta(g_{k}p_{k}+g_{m}p_{m})]+|h_{n}g_{n}|(\Delta x_{n})(\Delta p_{n})\geq|h_{k}g_{k}+h_{m}g_{m}|+|h_{n}g_{n}|$.
$\square$

\subsection{Proof of the product version of the van Loock-Furusawa inequalities
Eq. (\ref{eq:threeineq-1}) }

For $S_{I}$, we have the condition $h_{1}=-h_{2}=g_{1}=g_{2}=1$
and $h_{3}=0$. Using the result (\ref{eq:uvineq-1}), we see that
the states $\rho=\sum_{i}\eta_{i}^{(2)}\rho_{13}^{i}\rho_{2}^{i}$
and $\rho=\sum_{i}\eta_{i}^{(1)}\rho_{23}^{i}\rho_{1}^{i}$ satisfy
$S_{I}\geq2$, while the state $\rho=\sum_{i}\eta_{i}^{(3)}\rho_{12}^{i}\rho_{3}^{i}$
gives $S_{I}\geq0$. Similarly, we have $h_{2}=-h_{3}=g_{2}=g_{3}=1$
and $h_{1}=0$ for $S_{II}$. The states $\rho=\sum_{i}\eta_{i}^{(3)}\rho_{12}^{i}\rho_{3}^{i}$
and $\rho=\sum_{i}\eta_{i}^{(2)}\rho_{13}^{i}\rho_{2}^{i}$ satisfy
$S_{II}\geq2$ while the state $\rho=\sum_{i}\eta_{i}^{(1)}\rho_{23}^{i}\rho_{1}^{i}$
gives $S_{II}\geq0$. Lastly, the conditions $h_{1}=-h_{3}=g_{1}=g_{3}=1$
and $h_{2}=0$ for $S_{III}$ give $S_{III}\geq2$ for the states
$\rho=\sum_{i}\eta_{i}^{(3)}\rho_{12}^{i}\rho_{3}^{i}$ and $\rho=\sum_{i}\eta_{i}^{(1)}\rho_{23}^{i}\rho_{1}^{i}$,
and $S_{III}\geq0$ for $\rho=\sum_{i}\eta_{i}^{(2)}\rho_{13}^{i}\rho_{2}^{i}$.
$\square$

\subsection{Proof of Criterion (2)}

Consider any mixture of the form Eq. (\ref{eq:mixgen}). We can use
the result (\ref{eq:convexprod}) to write $S_{I}\geq P_{1}S_{I,1}+P_{2}S_{I,2}+P_{3}S_{I,3}\geq P_{1}S_{I,1}+P_{2}S_{I,2}$
where $S_{I,k}$ ($k=1,2,3$) is the value of $S_{I}$ predicted for
the component $k$ of the mixture. Now we know that the first two
states of the mixture satisfy the inequality, $S_{I}\geq2$. Hence,
for any mixture $S_{I}\geq2(P_{1}+P_{2})$. Similarly, $S_{II}\geq2(P_{2}+P_{3})$
and $S_{III}\geq2(P_{1}+P_{3})$. Then we see that since $\sum_{k=1}^{3}P_{k}=1$,
for any mixture it must be true that $S_{I}+S_{II}+S_{III}\geq4$.
$\square$

\subsection{Mixed bipartite entangled states that are fully tripartite inseparable}

Consider the mixed biseparable state of the type given by Shalm et
al. \cite{shalm-1} 
\begin{equation}
\rho_{BS}^{c}=\frac{1}{2}\rho_{12}\rho_{3}+\frac{1}{2}\rho_{23}\rho_{1}\label{eq:counter-2}
\end{equation}
This mixed state satisfies the van Loock- Furusawa criteria for full
tripartite inseparability but, being a mixture of biseparable states,
is not genuinely tripartite entangled. Here $\rho_{12}$ and $\rho_{23}$
are two-mode squeezed states defined by $\rho_{km}=|\psi_{km}\rangle\langle\psi_{km}|$
where $|\psi_{km}\rangle=(1-x^{2})^{1/2}\sum_{n=0}^{\infty}x^{n}|n\rangle_{k}|n\rangle_{m}$.
Here, $|n\rangle_{k}$ are the number states of mode $k$, $x=\tanh(r)$
and $r\geq0$ is the squeeze parameter that determines the amount
of two-mode squeezing (entanglement) between the modes $k$ and $m$.
The $\rho_{j}$ are single mode vacuum squeezed states, with squeeze
parameter denoted by $r$. The component $\rho_{12}\rho_{3}$ can
violate the inequality $B_{I}\geq4$, while $\rho_{23}\rho_{1}$ can
violate the inequality $B_{II}\geq4$. It is straightforward to show
on selecting $g_{1}=g_{3}=g$ that $\rho_{BS}$ can violate \emph{both}
inequalities. This demonstrates the full inseparability of the biseparable
mixture, by way of the van Loock-Furusawa inequalities. Unless one
can exclude mixed states, therefore, further criteria are needed to
detect genuine tripartite entanglement.

\subsection{Proof of Criterion (6)}

This follows from the result Eq. (\ref{eq:uvineq-1}). Using Eq. (\ref{eq:uvineq-1}),
we see that the bipartition given by $12-3$ implies $\Delta u\Delta v\geq|g_{3}h_{3}|+|h_{1}g_{1}+h_{2}g_{2}|,$
the bipartition $13-2$ implies $\Delta u\Delta v\geq|g_{2}h_{2}|+|h_{1}g_{1}+h_{3}g_{3}|,$
and the bipartition $23-1$ implies $\Delta u\Delta v\geq|g_{1}h_{1}|+|g_{2}h_{2}+h_{3}g_{3}|.$
Thus, we see that any mixture Eq. (\ref{eq:mixgen}) will imply Eq.
(\ref{eq:prodgen-1}). $\square$

\subsection{Proof of the relations Eq. (\ref{eq:VLFresult-4}) and Eq. (\ref{eq:prodst-1})
for EPR steering criteria}

For the special sort of bipartition $\{km,n\}_{s}$\emph{, only} system
$n$ is constrained to be a quantum state. Letting $u=h_{k}x_{k}+h_{m}x_{m}+h_{n}x_{n}$
and $v=g_{k}p_{k}+g_{m}p_{m}+g_{n}p_{n}$, we show that always 
\begin{eqnarray*}
[\Delta(h_{k}x_{k}+h_{m}x_{m}+h_{n}x_{n})]^{2}+[\Delta(g_{k}p_{k}+g_{m}p_{m}+g_{n}p_{n})]^{2}\\
\geq\sum_{i}\eta_{i}\{(h_{n}^{2}\Delta x_{n}^{2})_{i}+\Delta(h_{m}x_{m}+h_{k}x_{k})^{2}\\
+(g_{n}^{2}\Delta p_{n}^{2})_{i}+\Delta(g_{m}p_{m}+g_{k}p_{k})^{2}\},
\end{eqnarray*}
where we follow Ref. \cite{hoftake} and use that for a mixture, the
variance cannot be less than the average of the variance of the components.
Because the state of systems $k$ and $m$ is not assumed to be a
quantum state, there is only the assumption of non-negativity for
the associated variances. The single system $n$, however, is constrained
to be a quantum state, and therefore its moments satisfy the uncertainty
relation, which implies $(\Delta x_{n})^{2}+(\Delta p_{n})^{2}\geq2$.
Hence, if we assume that system $k,m$ cannot steer $n$, the following
inequality will hold: 
\begin{equation}
(\Delta u)^{2}+(\Delta v)^{2}\geq2|h_{n}g_{n}|.\label{eq:VLFresult-2}
\end{equation}
The product relation follows similarly. $\square$

\subsection{Proof of Criteria (1s) and (2s) }

We assume the hybrid LHS model associated with Eq. (\ref{eq:lhs-1})
is valid. Since then $B_{I}$ is the sum of two variances of a system
in a probabilistic mixture, we can write $B_{I}\geq P_{1}B_{I,1}+P_{2}B_{I,2}+B_{I,3}\geq P_{1}B_{I}+P_{2}B_{2}$
where $B_{I,n}$ denotes the prediction for $B_{I}$ given the system
is in the bipartition $\{km,n\}_{s}$. Now we know that the first
two states of the mixture satisfy the inequality, $B_{I}\geq2$. Hence,
for any mixture $B_{I}\geq2(P_{1}+P_{2})$. Similarly, $B_{II}\geq2(P_{2}+P_{3})$
and $B_{III}\geq2(P_{1}+P_{3})$. Then we see that since $\sum_{k=1}^{3}P_{k}=1$,
for any mixture it must be true that $B_{I}+B_{II}+B_{III}\geq4$.
Hence tripartite genuine steering is confirmed when this inequality
is violated. and Similarly, for the hybrid LHS model, $S_{I}\geq P_{1}S_{I,1}+P_{2}S_{I,2}+P_{3}S_{I,3}\geq P_{1}S_{I,1}+P_{2}S_{I,2}$
where $S_{I,n}$ ($n=1,2,3$) is the value of $S_{I}$ predicted given
the system is in the bipartition $\{km,n\}_{s}$. Now we know that
the first two states of the mixture satisfy the inequality, $S_{I}\geq1$.
Hence, for any mixture $S_{I}\geq P_{1}+P_{2}$. Also, $S_{II}\geq P_{2}+P_{3}$
and $S_{III}\geq P_{1}+P_{3}$, which implies $S_{I}+S_{II}+S_{III}\geq2$.
$\square$

\subsection{Proof of Criteria (3s) and (4s)}

\textbf{Proof:} First, we assume the system is described by the bipartition
$\{12,3\}_{st}$. Using Eq. (\ref{eq:VLFresult-4}) with $u=x_{1}-\frac{(x_{2}+x_{3})}{\sqrt{2}}$
and $v=p_{1}+\frac{(p_{2}+p_{3})}{\sqrt{2}}$, this gives the constraint
$(\Delta u)^{2}+(\Delta v)^{2}\geq1$. Similarly, the bipartition
$\{13,2\}_{st}$ gives $(\Delta u)^{2}+(\Delta v)^{2}\geq1$, and
the bipartition $\{23,1\}_{st}$ gives $(\Delta u)^{2}+(\Delta v)^{2}\geq2$.
Thus, all bipartitions satisfy $(\Delta u)^{2}+(\Delta v)^{2}\geq1$.
Using the result Eq. (\ref{eq:convesum}), for the system in a probabilisitc
mixture where moments are given as Eq. (\ref{eq:lhs-1}), we can say
that $(\Delta u)^{2}+(\Delta v)^{2}\geq1$. Thus, genuine tripartite
steering is confirmed if this inequality is violated. Using Eq. (\ref{eq:prodst-1})
for the bipartition $\{12,3\}_{st}$, it is also true that $\Delta u\Delta v\geq\frac{1}{2}$,
and similarly for bipartition $\{13,2\}_{st}$. For bipartition $\{23,1\}_{st}$
we find $\Delta u\Delta v\geq1$. Then again, for any mixture, using
Eq. (\ref{eq:convexprod}), we deduce Criterion (4s).\textcolor{red}{}
$\square$

\subsection{Optimising the Criterion (8)}

We describe the algorithm to compute the gains ($g,h$) used in the
Figures based on Criterion (8), for the GHZ and asymmetric and symmetric
EPR-type states. The variances $(\Delta u)^{2}$ and $(\Delta v)^{2}$
on the left-side of the inequality (37) can be expanded in terms of
covariance matrix elements of the inputs (following Ref. \cite{cvsig}),
which can then be computed for the relevant CV quantum state. We select
$h_{i}=h$ and $g_{i}=g$ for $ $for $i\geq2$. The choice of $g,h$
values was obtained by setting $\frac{d}{dh}(\Delta u)^{2}=0$ and
$\frac{d}{dg}(\Delta v)^{2}=0$. For the CV GHZ state, expanding we
have 

\begin{gather}
(\Delta u)^{2}=\frac{1}{N}[(N-1)^{2}h^{2}+2h(N-1)+1](\Delta x_{1}^{(in)})^{2}\nonumber \\
+\frac{(N-1)}{N}[h^{2}-2h+1](\Delta x_{2}^{(in)})^{2}\nonumber \\
(\Delta v)^{2}=\frac{1}{N}[(N-1)^{2}g^{2}+2g(N-1)+1](\Delta p_{1}^{(in)})^{2}\nonumber \\
+\frac{(N-1)}{N}[g^{2}-2g+1](\Delta p_{2}^{(in)})^{2}
\end{gather}
which gives on differentiation, the choice of 
\begin{gather}
h=-\frac{(\Delta x_{1}^{(in)})^{2}-(\Delta x_{2}^{(in)})^{2}}{(\Delta x_{2}^{(in)})^{2}+(N-1)(\Delta x_{1}^{(in)})^{2}}\nonumber \\
g=-\frac{(\Delta p_{1}^{(in)})^{2}-(\Delta p_{2}^{(in)})^{2}}{(\Delta p_{2}^{(in)})^{2}+(N-1)(\Delta p_{1}^{(in)})^{2}}
\end{gather}
Here, $(\Delta x_{1}^{(in)})^{2}=e^{2r}$ , $(\Delta x_{2}^{(in)})^{2}=e^{-2r}$
, $(\Delta p_{1}^{(in)})^{2}=e^{-2r}$, and $(\Delta p_{2}^{(in)})^{2}=e^{2r}$
are the variances for the two inputs to BS1, as depicted in Figure
9. The superscript $(in)$ denotes the input modes. For the $N=4$
configuration at large $r$, we see that $g=1$ and $h=-1/3$. In
general, for $g,h$ values satisfying $|gh|\leq1$, $gh<0$, $1-2gh\geq1$,
we see that the right-side of Criterion (8) reduces to $2[1+(N-3)gh]$.\textcolor{black}{{}
Identical procedures are used to obtain the gains for the asymmetric
EPR-type state I of Figure 10. They are given as:}

\begin{gather}
h=-\frac{(\Delta x_{1}^{(in)})^{2}-(\Delta x_{2}^{(in)})^{2}}{\sqrt{(N-1)}[(\Delta x_{2}^{(in)})^{2}+(\Delta x_{1}^{(in)})^{2}]}\nonumber \\
g=-\frac{(\Delta p_{1}^{(in)})^{2}-(\Delta p_{2}^{(in)})^{2}}{\sqrt{(N-1)}[(\Delta p_{2}^{(in)})^{2}+(\Delta p_{1}^{(in)})^{2}]}
\end{gather}
\textcolor{black}{For the $N=4$ configuration at large $r$, we see
that $g=1/\sqrt{3}$ and $h=-1/\sqrt{3}$.} For the symmetric EPR-type
state II of Figure 11, the analytical expressions depend on whether
the number of parties that are involved is even or odd. However, the
algorithm to compute these gains is otherwise identical.

\end{document}